\documentclass[sigconf]{acmart}

\usepackage{amsmath}
\usepackage{array}
\usepackage{makecell}
\usepackage{pifont}
\usepackage{graphicx}
\usepackage{booktabs} % For formal tables
\usepackage{multirow}
\usepackage{balance}
\usepackage{enumitem}
\usepackage{xcolor}
\usepackage{acmart-taps}

\AtBeginDocument{%
  \providecommand\BibTeX{{%
    \normalfont B\kern-0.5em{\scshape i\kern-0.25em b}\kern-0.8em\TeX}}}

\copyrightyear{2026}
\acmYear{2026}
\setcopyright{cc}
\setcctype{by-nc-nd}
\acmConference[CHI '26]{Proceedings of the 2026 CHI Conference on Human Factors in Computing Systems}{April 13--17, 2026}{Barcelona, Spain}
\acmBooktitle{Proceedings of the 2026 CHI Conference on Human Factors in Computing Systems (CHI '26), April 13--17, 2026, Barcelona, Spain}
\acmDOI{10.1145/3772318.3791273}
\acmISBN{979-8-4007-2278-3/2026/04}

\begin{document}

% \tableofcontents
% \clearpage

%  \title[[short title]{long title}

% \title[Expanding Communication Access]{Expanding Communication Access: Research Findings Incorporating AI as A Multi-Modal Approach to AAC}

\title[Giving Meaning to Movements]
{Giving Meaning to Movements: Challenges and Opportunities in Expanding Communication by Pairing Unaided AAC with Speech Generated Messages
}

% % Opportunities and Challenges in Expanding Communication Modalities for AAC Users by Pairing Unaided AAC with Speech Generated Messages

% \author{Syed Masum Billah}
% \authornote{author note}
% \affiliation{%
%   \institution{Pennsylvania State University}  
%   \city{University Park}
%   \state{PA}
%   \country{United States}
% }
% \email{sbillah@psu.edu}

% \shortauthors{Billah et al.}

\author[Kabir]{Imran Kabir}
% \authornote{author note}
\affiliation{%
  \institution{Pennsylvania State University}
  \city{University Park}
  \state{PA}
  \country{United States}
}
\email{ibk5106@psu.edu}

\author[Redmon]{Sharon Ann Redmon}
% \authornote{author note}
\affiliation{%
  \institution{Pennsylvania State University}
  \city{University Park}
  \state{PA}
  \country{United States}
}
\email{sar353@psu.edu}

\author[Elko]{Lynn R Elko}
% \authornote{author note}
\affiliation{%
  \institution{See CVI, Speak AAC}
  \city{Tamaqua}
  \state{PA}
  \country{United States}
}
\email{lynnelko88@gmail.com}

\author[Williams]{Kevin Williams}
% \authornote{author note}
\affiliation{%
  \institution{L. L. SLIM, LLC.}
  \city{Charlotte}
  \state{NC}
  \country{United States}
}
\email{llslim@gmail.com}

\author[Case]{Mitchell A Case}
% \authornote{author note}
\affiliation{%
  \institution{Pennsylvania State University}
  \city{University Park}
  \state{PA}
  \country{United States}
}
\email{mac8002@psu.edu}

\author[Sowers]{Dawn J Sowers}
\authornote{This project started while at Pennsylvania State University.}
\affiliation{%
  \institution{Florida State University}
  \city{Tallahassee}
  \state{FL}
  \country{United States}
}
\email{djs25d@fsu.edu}

\author[Wilkinson]{Krista M Wilkinson}
% \authornote{author note}
\affiliation{%
  \institution{Pennsylvania State University}
  \city{University Park}
  \state{PA}
  \country{United States}
}
\email{kmw22@psu.edu}

\author[Billah]{Syed Masum Billah}
% \authornote{author note}
\affiliation{%
  \institution{Pennsylvania State University}
  \city{University Park}
  \state{PA}
  \country{United States}
}
\email{sbillah@psu.edu}

\renewcommand{\shortauthors}{Kabir, Redmon, Elko, Williams, Case, Sowers, Wilkinson and Billah}

% A Dataset for Recognition of Crucial Objects in Navigation of Blind and Low-Vision People 

\begin{abstract}

Augmentative and Alternative Communication (AAC) technologies are categorized into two forms: aided AAC, which uses external devices like speech-generating systems to produce standardized output, and unaided AAC, which relies on body-based gestures for natural expression but requires shared understanding. We investigate how to combine these approaches to harness the speed and naturalness of unaided AAC while maintaining the intelligibility of aided AAC, a largely unexplored area for individuals with communication and motor impairments. Through 18 months of participatory design with AAC users, we identified key challenges and opportunities and developed AllyAAC, a wearable system with a wrist-worn IMU paired with a smartphone app. We evaluated AllyAAC in a field study with 14 participants and produced a dataset containing over 600,000 multimodal data points featuring atypical gestures—the first of its kind. Our findings reveal challenges in recognizing personalized, idiosyncratic gestures and demonstrate how to address them using Transformer-based large machine learning (ML) models with different pretraining strategies. In sum, we contribute design principles and a reference implementation for adaptive, personalized systems combining aided and unaided AAC.

% Augmentative and Alternative Communication (AAC) technologies are categorized into two forms: aided AAC, which uses external devices like speech-generating systems to produce standardized output; and unaided AAC, which relies on body-based gestures for natural expression but depends on shared understanding. In this paper, we investigate how to combine these approaches to harness the speed and naturalness of unaided AAC while maintaining the intelligibility of aided AAC, a largely unexplored area for individuals with communication and motor impairments. Through 18 months of participatory design with AAC users, we identified key challenges and opportunities and iteratively developed AllyAAC, a wearable system with a wrist-worn IMU sensor paired with a smartphone app. We evaluated AllyAAC in a field study with 14 participants in a social context and produced a dataset containing over 600,000 multimodal data points featuring atypical gestures paired with IMU data—the first of its kind. Our findings reveal challenges in recognizing personalized idiosyncratic gestures and demonstrate how to address them using large machine learning (ML) models based on the Transformer architecture with different pretraining strategies. Our results also indicate that different pretraining strategies work best for different motor-impaired individuals. In sum, we contribute design principles for adaptive, personalized systems that combine aided and unaided AAC and provide a reference implementation to inspire AAC, HCI, and ML researchers.
\end{abstract}
%%

% https://dl.acm.org/ccs#

\begin{CCSXML}
<ccs2012>
   <concept>
       <concept_id>10003120.10003121.10003124</concept_id>
       <concept_desc>Human-centered computing~Interaction paradigms</concept_desc>
       <concept_significance>300</concept_significance>
       </concept>
   <concept>
       <concept_id>10003120.10011738.10011775</concept_id>
       <concept_desc>Human-centered computing~Accessibility technologies</concept_desc>
       <concept_significance>500</concept_significance>
       </concept>
   <concept>
       <concept_id>10003120.10003138.10003141.10010898</concept_id>
       <concept_desc>Human-centered computing~Mobile devices</concept_desc>
       <concept_significance>300</concept_significance>
       </concept>
   <concept>
       <concept_id>10003120.10003123.10010860.10010858</concept_id>
       <concept_desc>Human-centered computing~User interface design</concept_desc>
       <concept_significance>500</concept_significance>
       </concept>
 </ccs2012>
\end{CCSXML}

\ccsdesc[300]{Human-centered computing~Interaction paradigms}
\ccsdesc[500]{Human-centered computing~Accessibility technologies}
\ccsdesc[300]{Human-centered computing~Mobile devices}
\ccsdesc[500]{Human-centered computing~User interface design}

\keywords{Augmentative and Alternative Communication, AAC, Speech Impairments; Motor Impairments; Input Techniques, Body-based communication, Gestures; Machine Learning, Personalization, Datasets; IMU sensors, Wearable sensors; Participatory Design; Assistive Technologies.}

% \begin{teaserfigure}
%     \centering    
%         \includegraphics[width=.99\linewidth]{figures/plots_frames/image_plots/video-3-segment-2.pdf}                
%         \caption{
%         An illustration of how the perception of similarity between two consecutive keyframes (image) in a video differs for humans and AI models. (Top) shows a video clip containing 7 keyframes. (Bottom) shows the normalized inter-frame similarity scores, as perceived by sighted humans (green line), by a visual-question answering (VQA) model (red line), and by an image-based feature extractor (blue line). Notice that humans overlook minor changes in two consecutive frames and consider them similar,  as the mostly flat line indicates. The image feature-based similarity score correlates to humans' perception. 
%     }
%   \Description{}
%   \label{fig:teaser}
% \end{teaserfigure}

\maketitle
%%%%%%%%%%%%%%%%%%%%%%%%%%%%%%%%%
\section{Introduction}
\label{sec:introduction}
Augmentative and alternative communication (AAC) technologies are broadly categorized into \textit{aided} and \textit{unaided} forms. Aided AAC is the most widely used technology for individuals with communication disorders. It requires external tools or devices such as paper-based communication boards or high-tech systems such as speech-generating devices on computers or tablets~\cite{beukelman2020augmentative}. Unaided AAC, in contrast, includes communication methods that rely solely on the user's body, such as gestures and facial expressions. 
Familiar communication partners (e.g., aides, family members) can interpret these gestures and expressions with relative ease. Machine learning models, such as multi-class classifiers, can also recognize them automatically when trained on an individual's gestures captured through a camera or IMU (inertial measurement unit) sensor data, which are built into consumer devices like smartphones or smartwatches.
Traditionally, aided and unaided AAC have been considered mutually exclusive at any given moment.

Aided AAC provides the clear benefit of standardized and intelligible messages. The output is spoken with synthesized speech by the device or is shown as text on its display. As such, communication partners do not need to infer intent from body movements. This makes aided AAC especially valuable when interacting with unfamiliar communication partners. However, it places the burden on users to efficiently and accurately compose messages using the interface rendered on low- or high-tech displays, which they access through finger pointing, head pointers, switch activation, or eye gaze. Many individuals with communication disorders also experience co-occurring disabilities such as vision impairment or motor difficulties, and they may find it difficult to interact with these interfaces because they are too small, crowded, or require fine hand–eye coordination. They may also need to mount their AAC devices on wheelchairs to accommodate motor challenges~\cite{sennott2019aac}. These restrict the spontaneity of their communication.

Unaided AAC poses a different set of trade-offs. Users can produce body-based gestures at any time, often more quickly and naturally than when navigating a device. For example, a shoulder shrug can signal ``I don't know,'' a head tilt can mean ``over there,'' and a sweeping hand can mean ``you first.'' However, unaided AAC depends on shared understanding. Idiosyncratic movements, while effective with familiar partners, often fail with unfamiliar ones~\cite{blackstone2021children, wilkinson2025consideration}. Commercial gesture-recognition systems, typically built on able-bodied datasets, rarely capture the movement repertoire of people with motor impairments. As a result, the potential speed and naturalness of unaided AAC are offset by its limited reach across communication partners.

In this paper, we investigate how to harness the benefits of both aided and unaided AAC. More specifically, how unaided body-based communicative movements can serve as expressive input to traditional aided AAC devices, where the former provides the speed and naturalness of gesture and the latter provides the standardized output of speech or text.
Our research questions are as follows:  
\aptLtoX[graphic=no,type=html]{
\begin{itemize}
    \item [\textbf{RQ1:}] \textit{What are the opportunities and challenges of combining aided and unaided AAC?}  
    \item [\textbf{RQ2:}] \textit{How do users perceive this combination and in what contexts do they intend to use it (or avoid it)?}
\end{itemize}
}{
\begin{itemize}
    \item [\textbf{RQ1:}] \textit{What are the opportunities and challenges of combining aided and unaided AAC?}  
    \item [\textbf{RQ2:}] \textit{How do users perceive this combination and in what contexts do they intend to use it (or avoid it)?}
\end{itemize}
}

To address these questions, we conducted our research in the following phases.
\paragraph{\textbf{Participatory Design (PD) Sessions}}
We adopted participatory design (PD) for more than 18 months. Our research team includes domain experts (speech-language pathologists, assistive technology developers), system designers, and three community advisors with varying levels of motor and vision impairments whose lived experience set priorities (Section~\ref{subsec:pd_session}).
Through PD sessions, we identified several opportunities for unaided AAC, most notably, to convey quick messages, to communicate with partners who are in non-line-of-sight, to communicate emotion, and to quickly create rapport with unfamiliar communication partners (Section~\ref{subsec:opportunities_pd}). We also identified several key challenges, such as difficulty in recognizing personalized gestures with high accuracy, issues with synchronized data recording, clutching to prevent false activations, automatic calibration for sensor orientation changes, and adapting to users' personal context-dependent gesture repertoires (Section~\ref{subsec:challenges_pd}).

\paragraph{\textbf{Iteratively Developing AllyAAC App}} 
Based on findings from the PD sessions, we iteratively developed AllyAAC (Section~\ref{sec:system}). We first sent wearable IMU sensors to community advisors by mail, then released AllyAAC as an Android app (see Fig.~\ref{fig:initial_setup}) that they could download and update over the air. In each iteration, the community advisors used AllyAAC for several days, and we improved the prototype based on their feedback.
AllyAAC allows AAC users and their aides to record personalized gestures and annotate them interactively from synchronized video and IMU data. It contains a simple, rule-based gesture recognizer (our baseline model) that users can configure interactively (Section~\ref{subsec:baseline_model}). It can also load personalized large models trained on their data (Section~\ref{subsec:large_model}).
     
\paragraph{ \textbf{A Field Study in Social Contexts}} 
We conducted a field study with 14 participants in a social context (at the ATIA'25 Conference) to evaluate the feasibility, usability, and recognition performance of body-based communicative gestures (Section~\ref{sec:qual_eval}).
We observed how participants interacted with the system to communicate with others in a crowded space.
We created a dataset during this study (with participants' consent) that contained over 600,000 (0.6M) multimodal data points (videos of participants making different gestures and associated IMU sensor logs). To our knowledge, this is the first dataset of this kind to feature atypical, non-normative gestures paired with IMU data. 
This study confirmed that use cases from the PD sessions were indeed desirable and much sought after. 
It also revealed additional challenges for combining unaided body-based input with aided AAC output.
For instance, AAC users with motor impairments perform atypical gestures that can differ even for the same intent, which led to poor accuracy of the baseline model ($F_1$-score under 60\%). This result highlights the trade-off between the ease of model creation, the complexity of the model, and the model performance.

\paragraph{ \textbf{Design of Large Personalized Gesture Recognition Models}}
We designed three large machine learning (ML) models based on the Transformer architecture~\cite{vaswani2017attention} to improve recognition performance over the baseline model (Section~\ref{subsec:large_model}). Since large models require many manually annotated data points, which our participants stated as ``not feasible,'' we developed a semi-automatic annotation pipeline (Section~\ref{subsec:gesture_annotation}). In this pipeline, we used an off-the-shelf library (MediaPipe~\cite{zhang2020mediapipe}) to segment video clips that contain multiple instances of the same gesture, separated by idle periods. This pipeline reduced annotation time by 66\% compared to manual annotation for the same gestures (Section~\ref{subsec:eval_annotation_pipeline}).

\paragraph{ \textbf{Pretraining Strategies for Large Models}} 
We also investigated various pretraining strategies, particularly contrastive learning~\cite{tang2020exploring}, masked reconstruction~\cite{haresamudram2020masked}, and CPC~\cite{haresamudram2021contrastive} (which predicts latent representations of future timesteps), to further reduce the need for manually annotated data. We discovered that different pretraining strategies work best for different participants (Section~\ref{subsec:finding_pretraining}). For example, CPC works best for gestures with longer trajectories, as it captures long-term temporal characteristics. These large models improved recognition performance substantially ($F_1$-score over 85\%), representing a 28\% gain over the baseline model (Section~\ref{subsec:finding_classification}).

\paragraph{ \textbf{Human Evaluation of Large Models}} 
To ensure that the performance of our large models did not drift while deployed, we developed an evaluation tool (Fig.~\ref{fig:eval_ui}) that allows human annotators (N=6) to watch videos of a participant making different gestures along with the model predictions, so they could mark whether the predicted gesture accurately reflects the participant's intent.
Using this tool, we found that the mean precision score of the large models was over 0.85, suggesting good alignment between model predictions and human judgment (Section~\ref{subsec:human_eval}). The inter-annotator agreement was high (AC1~\cite{gwet2008computing} = 0.92), which suggests good alignment across different human judgments. Our community advisors currently use these models personalized for their gestures.

We summarize our contributions as follows:
\aptLtoX[graphic=no,type=html]{
\begin{itemize}
    \item Design principles for combining body-based communicative gestures with existing aided AAC, derived from 18 months of participatory design with AAC users.
    \item Identification of key challenges in recognizing personalized body-based communicative gestures captured through IMU sensors with machine learning models.    
    \item A multimodal dataset of 600,000 data points featuring non-normative gestures paired with IMU data.
    \item A semi-automatic annotation pipeline that speeds up the annotation process by $3\times$.
    \item Three personalized Transformer-based gesture recognition models with pretraining strategies and an offline human evaluation tool.
\end{itemize}
}{
\begin{itemize}[nosep]
    \item Design principles for combining body-based communicative gestures with existing aided AAC, derived from 18 months of participatory design with AAC users.
    \item Identification of key challenges in recognizing personalized body-based communicative gestures captured through IMU sensors with machine learning models.    
    \item A multimodal dataset of 600,000 data points featuring non-normative gestures paired with IMU data.
    \item A semi-automatic annotation pipeline that speeds up the annotation process by $3\times$.
    \item Three personalized Transformer-based gesture recognition models with pretraining strategies and an offline human evaluation tool.
\end{itemize}
}

\section{Background and Related Work}
\label{sec:related_work}
Our research addresses the challenge of combining aided and unaided AAC through gesture recognition. We review AAC systems and their limitations, the role of communication partners, prior work in gesture recognition for AAC, approaches to gesture-to-speech production, and machine learning techniques for recognizing personalized gestures.

\subsection{Augmentative and Alternative Communication (AAC)}
AAC technology includes a wide range of tools and techniques used to support or replace speech for individuals with communication challenges~\cite{beukelman2020augmentative}. Aided AAC systems include low-tech solutions such as communication boards and high-tech tools like speech-generating devices (SGDs) or tablet-based apps. These devices typically rely on visual-graphic interfaces with layered pages and symbolic vocabularies, accessed via direct touch, switch scanning, or eye gaze~\cite{sowers2023demands}.

While aided AAC can provide consistent message output, it places high demands on motor precision, visual acuity, and environmental stability. Glare, water exposure, or the need for specialized mounts can limit use outdoors or during movement~\cite{wilkinson2022judicious}. These limitations are especially pronounced for users with both motor and visual impairments.

Unaided AAC refers to body-based forms of communication—\allowbreak gestures, facial expressions, and other natural movements that require no external device. These can be efficient and expressive, but only when communication partners understand the user's individual repertoire~\cite{blackstone2021children}.

\subsubsection{AAC and Co-Occurring Motor and Visual Impairments}
The intersection of motor and visual impairments further complicates AAC use. Motor differences may limit direct access to small visual targets, while visual field deficits or oculomotor conditions such as nystagmus or strabismus affect symbol scanning and selection~\cite{sowers2023demands, wilkinson2004contributions}. Cortical visual impairment (CVI), in particular, disrupts functional visual processing and is often comorbid with motor disabilities~\cite{chang2024special}.

These combined impairments make both aided and unaided AAC fragile in real-world use. For instance, aided systems may require searching across cluttered grids, which causes fatigue or misselection~\cite{van2025relation, wilkinson2022judicious}, while unaided gestures may be overlooked or misinterpreted by unfamiliar partners. In practice, many families fall back on body-based gestures, even when aided AAC is technically available~\cite{blackstone2021children}. By converting personalized, embodied gestures into universally intelligible speech, our approach addresses a longstanding tension in AAC: balancing expressiveness and legibility, independence and reliability.

\subsubsection{Communication Partners and Interpretation Burden}
AAC is not just a tool—it is a socially co-constructed process. Effective communication often depends as much on the partner as on the AAC system itself~\cite{clarke2013aac, kent2015effects}. Familiar partners may become fluent in interpreting unaided AAC, but unfamiliar partners, including caregivers, transportation staff, or coworkers, often cannot. This creates a dependency that limits participation in spontaneous or public interactions. Given high turnover rates among personal care attendants (often exceeding 77\%) and the increasing engagement of AAC users in community life, there is a pressing need for systems that reduce reliance on partner familiarity~\cite{friedman2021impact, rollison2023evaluation}. Our system addresses this by shifting the interpretive burden from the partner to the algorithm.

\subsection{Gesture Recognition in AAC}
Researchers have explored gesture recognition as an input method for AAC~\cite{higginbotham2007access, boster2017you, koch2019new}. Early efforts relied on predefined gesture vocabularies and assumed normative movement trajectories, which made them unsuitable for users with non-standard motor patterns. More recent work suggests that individualized movement patterns—idiosyncratic but consistent—can be learned by machine learning models and used as meaningful input~\cite{wilkinson2025consideration}. Parallel work has applied large language models to AAC to predict communicative intent~\cite{valencia2023less}, but users consistently emphasize the need for control, personalization, and independence.

Gesture-based interaction systems typically distinguish between touch gestures (taps, swipes on touchscreens) and motion gestures (movements captured by IMU sensors or cameras). Touch gestures follow location-specific conventions for sighted users and location-agnostic patterns for blind users~\cite{slide_rule}. Motion gestures occur when users move a device (e.g., tilt, flip)~\cite{ruiz_user_defined_gesture} or move their hands in midair while wearing sensors. The semantics of motion gestures are often predefined (e.g., finger counting~\cite{ehtesham_abacus}) and used for navigation, selection, and activation~\cite{koutsabasis2019empirical, morris2014reducing}. 

Unaided AAC largely falls under midair gestures, but unlike most midair gesture systems, where semantics are predefined, communicative gestures for AAC users with motor impairments are deeply personalized, shaped by individual motor abilities, and grounded in daily life.

Another challenge in realizing unaided AAC is the mobility bias in existing datasets. Most public datasets have been collected from able-bodied participants performing predefined gestures and activities. Popular datasets such as UCI-HAR~\cite{uci_har}, OPPORTUNITY~\cite{opportunity_challenge}, PAMAP2~\cite{pamap2}, UTD-MHAD~\cite{utd_mhad}, and Berkeley MHAD~\cite{ucb_mhad} contain multimodal recordings (e.g., IMU, video, depth) of generic actions such as walking, running, waving, or clapping. As a result, models trained on these corpora often fail to generalize to atypical, idiosyncratic movements produced by individuals with motor impairments~\cite{nair2023dataset,torring2024validation} because they assume consistent, repeatable trajectories~\cite{kamikubo2022data,torring2024validation}.

Our contribution extends this literature by introducing a user-customized, gesture-driven interface and evaluating its feasibility with motor-impaired AAC users in real-world conditions.

\subsection{Gesture-to-Speech Production}
Prior work in gesture-to-speech production has predominantly focused on sign language recognition~\cite{koller2016deep,forster2014extensions,bragg2019sign}, which handles standardized vocabularies captured through high-quality video. Sign languages are characterized by phonological features (handshape, location, movement)~\cite{stokoe2005sign} and non-manual markers including facial expressions, eyebrows, mouth, head, shoulders, and eye gaze~\cite{wilbur2013phonological}. Recent advances have explored embedding-based evaluation to capture these multimodal aspects, including spatial grammar~\cite{imai2025silverscore}. 

Our work differs from sign language recognition in fundamental ways. Unlike sign languages, which have well-defined semantics, standardized vocabularies, and large-scale training datasets (e.g., PHOENIX-14T~\cite{camgoz2018neural}, CSL-Daily~\cite{zhou2021improving}), the gestures of motor-impaired users often lack a common vocabulary, and datasets containing such gestures are scarce. We offer a dataset that features synchronized IMU and video from AAC users with motor impairments to address this gap.

\subsection{Architectures of Machine Learning Models for Gesture Recognition}
Early work on gesture recognition and human activity recognition based on sensor data used hand-crafted statistical and spectral features from accelerometers and gyroscopes fed to shallow classifiers~\cite{bao2004activity,plotz2011feature}. End-to-end deep learning models later replaced manual feature engineering. Convolutional neural networks (CNNs) improved recognition by learning local motion patterns~\cite{ha2015multi,hammerla2016deep}, while hybrid CNN–RNN models captured longer-range temporal dependencies and inter-sensor relationships~\cite{zeng2018understanding}. Attention mechanisms then improved performance by emphasizing salient temporal and modality features~\cite{murahari2018attention,ma2019attnsense}. Transformers extended this idea by modeling global dependencies directly~\cite{vaswani2017attention}. These successes motivated us to design Transformer-based, large models that can run in real-time on AAC devices.

However, the key challenge with Transformer-based models is that they require large datasets~\cite{dosovitskiy2020image}. At the same time, personalized modeling is needed to accommodate the heterogeneity of motor-impaired users' motion. Prior work has noted that per-user variations demand adaptive or user-specific models~\cite{buffelli2021attention}. Training a separate model per user or fine-tuning to individual motor profiles is a potential solution, but data scarcity makes this challenging. 

\subsubsection{Transfer and Data Augmentation}
Recent advances in IMU-based activity recognition aim to reduce dependence on large labeled datasets through self-supervised and few-shot transfer learning. For example, \textit{SimCLR}~\cite{tang2020exploring} applies contrastive learning to maximize agreement between augmented views of the same signal~\cite{chen2020simple}. Haresamudram et al. propose masked reconstruction and contrastive predictive coding (CPC) to learn representations of motion sequences without labels~\cite{haresamudram2020masked,haresamudram2021contrastive}, improving downstream classification when labeled data are sparse. Since labeled datasets of idiosyncratic, motor-impaired communicative gestures are scarce and inter-user variability is high, we explore self-supervised pretraining on unlabeled IMU data, followed by supervised personalization with a small amount of annotated data. This strategy allows us to adapt models to individual users while addressing the data scarcity inherent to this domain.
\section{Understanding Opportunities and Challenges in Combining Aided \& Unaided AAC}
\label{sec:opp_and_challenges}

\subsection{Participatory Design (PD) Sessions}
\label{subsec:pd_session}
To address our research questions, we adopted a participatory design approach over 18 months (IRB approved). The team included domain experts, designers, developers, and three community advisors who are AAC users with motor or vision impairments. We refer to the three advisors as A1–A3. Their lived experience led our design priorities.

\paragraph{\textbf{Community Advisors}}
Advisors had upper- and lower-body motor impairments of varying presentation and relied on aided AAC in daily life. They also represented different personal and professional backgrounds. A1 is a freelance contractor and public speaker with experience spanning paper-based systems to tablet AAC with direct selection. A2 is a snow skier with cortical visual impairment who uses a wheelchair with an AAC device mounted. The mother of A2 frequently represented A2, sharing her perspective as the primary caregiver. A3 is a college student and a full-time wheelchair user. Their differences brought complementary perspectives on how communication needs arise in work, recreation, and education.

\paragraph{\textbf{Methods}}
The group met monthly over Zoom for 18 months, and community advisors communicated via a combination of aided and unaided AAC, as well as verbally to the extent of their capacity. Each session began with an open discussion, where the advisors discussed communication breakdowns and opportunities drawn from their daily routines. The role of the researchers was to probe, document, and brainstorm with advisors on how gesture-based support might intervene. Between sessions, researchers created prototypes that embodied the ideas of the advisors. They shipped IMU sensors via postal mail (once) and accompanying software (described in the following section) over the air for at-home and everyday use. 

The advisors tested prototypes in everyday settings for 1–2 weeks, documenting successes, breakdowns, and friction. Their reflections directly guided subsequent iterations. This iterative cycle (ideation, prototyping, real-world testing, reflection) ensured that design decisions were grounded in the lived practices of AAC users. Advisors shaped not only interface labels (e.g., replacing ``record data'' with ``show my gesture''), but also deeper system mechanics, such as when feedback should be delivered and how to suspend recognition.

After a major iteration, the research team invited community advisors to meet in person to observe first-hand how the prototype worked in a controlled lab environment. In these sessions, the team also recorded gesture data for the next revision. 

% This model reflects the principles of mutual learning and sharing of power in PD. The research team gained insight into embodied communication practices, while the advisors influenced technical directions and vocabulary in ways that reflected their expertise. 
One limitation of PD sessions was that much of the collaboration took place through remote meetings and distributed prototypes, which may not fully capture the dynamics of in-person co-design. Nevertheless, this structure allowed advisors to remain sustained partners over an extended period despite geographic separation.

\subsection{Findings: Opportunities for Combined AAC}
\label{subsec:opportunities_pd}
PD sessions revealed opportunities and challenges in combining aided and unaided AAC. These sessions also helped the team identify design specifications that reflect lived experience, technical feasibility, and expectations of how such technology should work in everyday contexts. 

\subsubsection{Composing Quick Messages}
Advisors often needed to communicate quick needs, such as ``I need a drink'' or ``I need to use the bathroom.'' They explained that composing such messages on an AAC device was slow and required sustained attention to the screen. In these moments, they preferred unaided gestures, such as a wrist flip, that could trigger speech output immediately.

\subsubsection{Communication Afar or Near}
The advisors described scenarios where they needed to communicate with others at a distance, particularly when partners were not in line of sight (e.g., playgrounds, skiing, surfing, or crowded events). For example, A2 wore an IMU sensor while watching a baseball game with her family. The sensor was connected via Bluetooth to an Android phone used by A2's mother, where our prototype ran. When A2's mother was not in A2's line of sight, she could still hear A2's excitement expressed through different hand movements and interpreted by our prototype. She described the moment as emotionally powerful, as it allowed her to share in A2's joy from afar.

A1 shared another personal anecdote involving his newborn. Before losing his natural voice, he voice-banked greetings such as ``Hi [child's name].'' He mapped these to wave gestures and connected his tablet (running our prototype) to a Bluetooth speaker placed on the crib. When he waved his hand, making eye contact with his child—without looking at the AAC device to compose the greeting—the prototype read out the greeting in his banked, authentic voice. A1 mentioned that the prototype allowed him to call his child in a way that felt natural and that preserved his identity.

\subsubsection{Communicating Emotions}
Advisors often used distinct movements to express emotion (e.g., ``I don’t like that,'' ``I’m happy'') or urgency (e.g., repeating a gesture rapidly to mean ``I need it NOW''). Familiar partners usually understood these cues, but unfamiliar partners sometimes misinterpreted them as agitation or meltdown. Body-based input paired with audible AAC output was therefore seen as essential to communicate affect clearly.

\subsubsection{Quick Onboarding with New Partners}
Gesture-to-speech was also viewed as valuable in clinical and caregiving contexts. A3 noted that new clinicians and aides often needed time to learn their personal gestures, creating delays and frustration. A system capable of directly interpreting their movements could reduce the onboarding time and support smoother interactions with unfamiliar partners. 

\subsubsection{Interacting with the Wheelchair}
Some advisors expressed interest in using gestures for wheelchair control, for example, a rightward hand swipe to rotate right and a raised hand to stop. They stressed, however, that such controls would need safety constraints and gating mechanisms to prevent accidental activation.

\subsection{Findings: Challenges in Combined AAC}
\label{subsec:challenges_pd}
\subsubsection{Movement Recording with Synchronized Video and Sensor Data}
The advisors emphasized that the data recording must be simple, accessible to the user or an aide, and require minimal navigation. The early prototypes captured only raw IMU streams, which were opaque without context. The advisors reported that it was difficult to locate gesture instances by viewing the sensor trace alone; adding contextual video made the recordings interpretable. We first added video capture via an external camera, but advisors found manual synchronization with IMU data to be burdensome and error-prone. We therefore integrated synchronized in-app video and IMU recording, which resolved alignment issues and streamlined annotation. Viewing motion alongside the sensor trace enabled advisors and aides to annotate data independently.

\begin{figure*}[t!]
    \centering
    \includegraphics[width=0.98\linewidth]{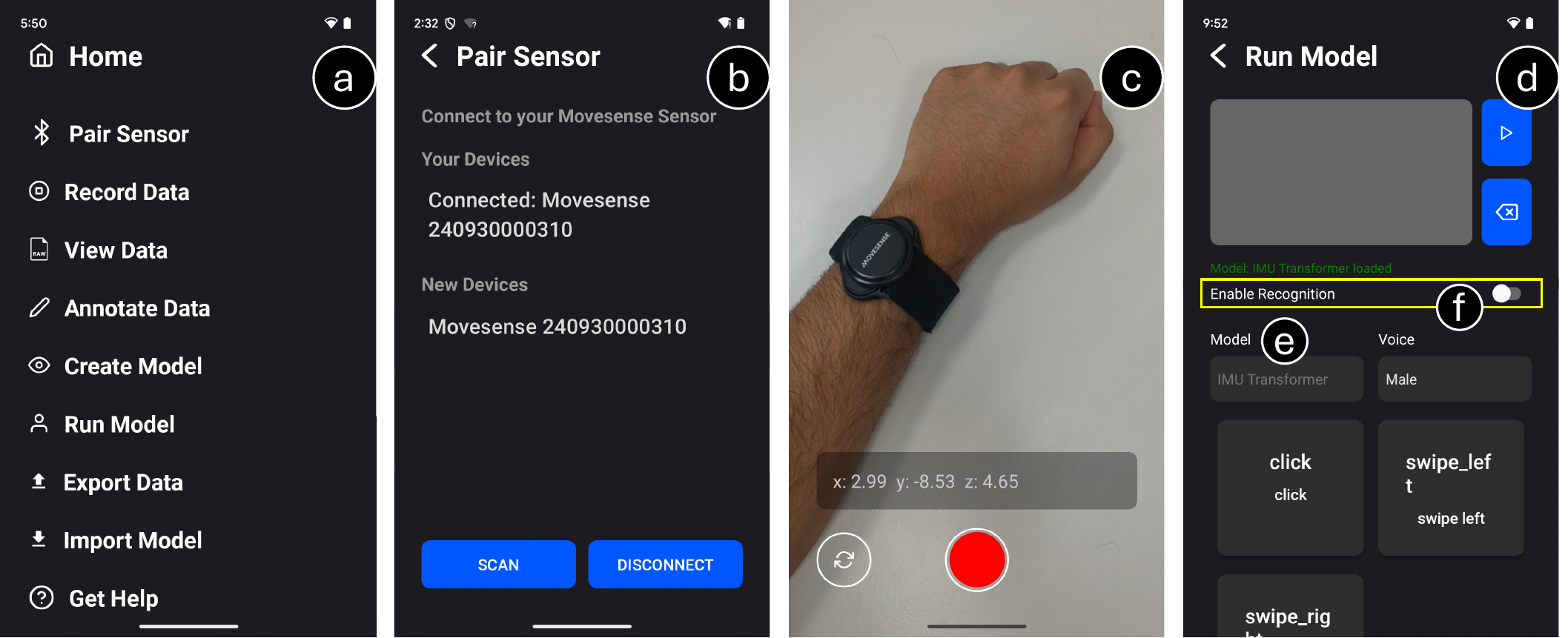}
    \Description{A composite image showing four screenshots of the AllyAAC mobile app interface. Panel (a) displays the home screen with app functionality options. Panel (b) shows the sensor pairing window with sections for paired devices and new devices. Panel (c) displays the data recording interface alongside a photo of an IMU sensor worn on a user's wrist. Panel (d) shows the gesture recognition testing window with a model selection dropdown (e), a toggle button for enabling/disabling recognition (f), and text-to-speech output functionality.}
    \caption{(a) The home screen of AllyAAC app showing all the available functionality of the app. (b) The window for pairing the sensor to the app. Paired sensors appear under ``Your Devices,'' while new sensors appear under ``New Devices.'' (c) The data recording window, along with how the IMU sensor is worn on the wrist. (d) The “Run Model” window, where the user can select a model from the dropdown (e) and test real-time gesture recognition. The interface also provides a toggle button (clutching mechanism) for enabling and disabling recognition (f) and speaks the message associated with each gesture using a text-to-speech (TTS) engine.}
    \label{fig:initial_setup}
\end{figure*}

\subsubsection{Clutching and Error Recovery}
The advisors described situations where everyday hand movements could accidentally trigger recognition. During conversation, their hands are often in motion—gesturing, shifting position, or manipulating objects—and these non-communicative movements should not produce speech output. To prevent false positives, they asked for a ``clutching'' mechanism to pause recognition temporarily. They proposed both on-screen and wheelchair-mounted buttons to suspend/resume recognition on demand. Alongside clutching, they requested a corrective gesture to cancel an unintended activation, analogous to saying ``never mind.'' This would allow immediate error recovery without interrupting conversational flow.

\subsubsection{Sensor Calibration and Orientation}
Calibrating the sensor emerged as another recurring challenge. Advisors disliked magnetometer calibration routines, describing them as disruptive and unrealistic for daily use. They also noted that recognition accuracy degraded when sensor orientation shifted, for example, when a wristband rotated. They requested automatic calibration and background orientation compensation so that the system could adapt without user intervention. Hardware placement was also a point of divergence. A2 preferred a wheelchair-mounted sensor for comfort, while A1 wanted two wrist-worn sensors to switch between hands depending on context. For example, when typing with the right hand, A1 could still use the left hand to make quick requests such as “Please bring me a glass of water.” These differences underscore the need for flexible sensor placements and personalized data fusion strategies.

\subsubsection{Personalization and Context}
The advisors stressed that gestures are personal and context-dependent. A single gesture, such as a wave, could carry multiple meanings depending on the setting (e.g., as a greeting at home or a refusal in a hospital). They wanted the system to account for this variability, learn their personal repertoire in context, and adapt over time.

In summary, the findings of the PD sessions highlight both the opportunities and challenges of combining aided and unaided AAC. They address RQ1 by identifying technical and interactional challenges such as synchronization, calibration, error recovery, and context dependence. They address RQ2 by showing how users perceive gesture-to-speech as valuable for quick needs, spontaneous interaction, non-line-of-sight communication, and eye-to-eye exchanges, as well as for reducing onboarding time with new caregivers.

\section{AllyAAC: A Combined Aided and Unaided AAC}
\label{sec:system}
\label{sec:workflow}
Based on the findings of our participatory design sessions, we developed AllyAAC, a system that enables real-time recognition of personalized body-based gestures. It is designed for AAC users who also have motor impairments. We describe the system workflow, from initial setup and gesture recording to annotation, model training, and deployment.

As noted in prior sections, data scarcity and the need for per-user personalization are key challenges in this domain. To address these challenges, AllyAAC facilitates data collection and annotation by end users or their aides, who can also configure a simple baseline model that uses only a few annotated examples. The full pipeline includes:
\aptLtoX[graphic=no,type=html]{
\begin{enumerate}
    \item Initial setup: wearing the IMU sensor and connecting it to the smartphone app;
    \item Recording template gestures;
    \item Semi-automatic gesture annotation using synchronized video and sensor data;
    \item Building the rule-based baseline model;
    \item Building the large model;
    \item Training and deploying personalized models for real-time recognition.
\end{enumerate}
}{
\begin{enumerate}
    \item Initial setup: wearing the IMU sensor and connecting it to the smartphone app;
    \item Recording template gestures;
    \item Semi-automatic gesture annotation using synchronized video and sensor data;
    \item Building the rule-based baseline model;
    \item Building the large model;
    \item Training and deploying personalized models for real-time recognition.
\end{enumerate}
}

\begin{figure*}[t!]
    \includegraphics[width=0.98\linewidth]{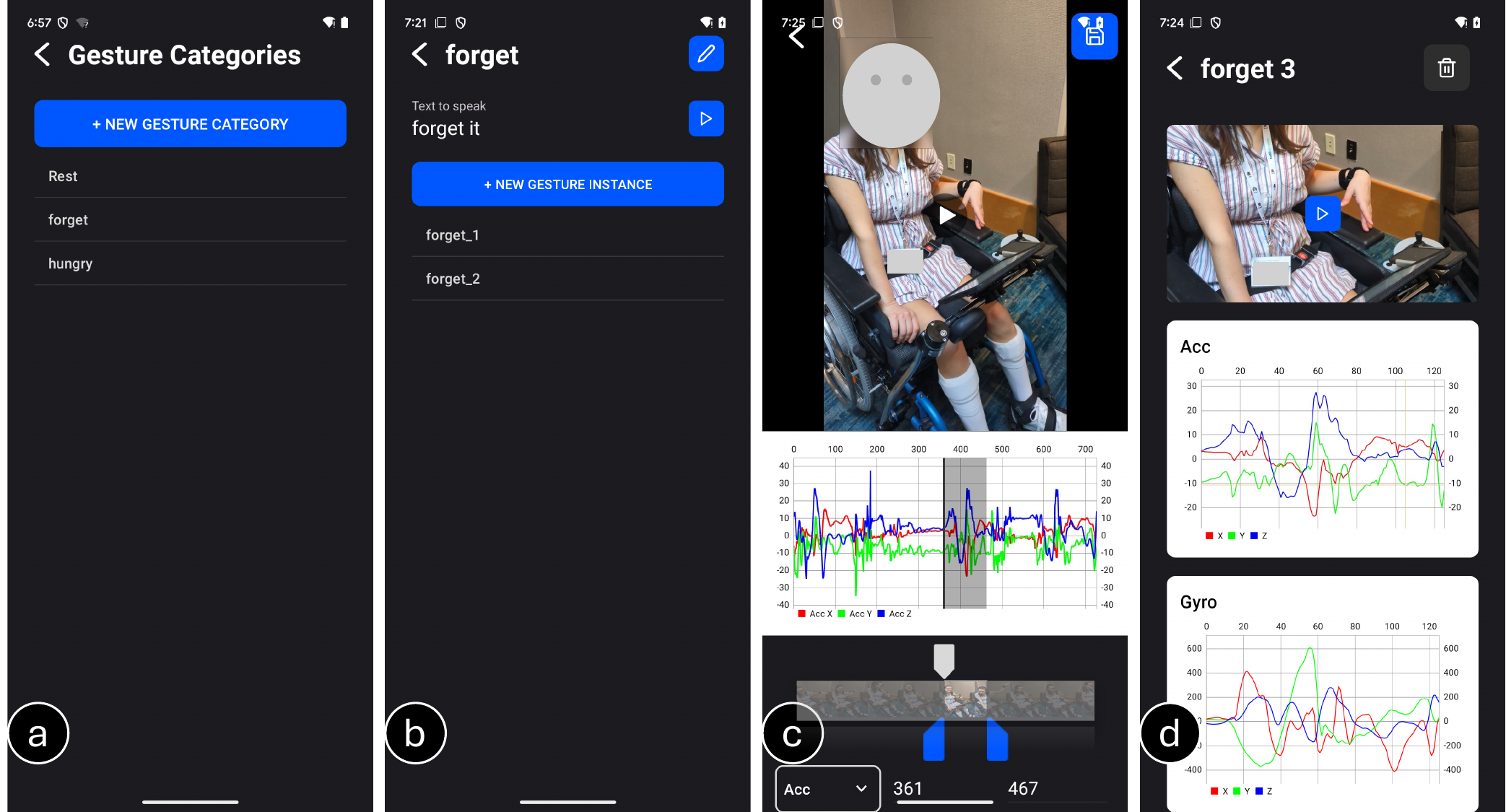}
    \Description{A composite image showing four screenshots of the gesture annotation workflow in the AllyAAC app. Panel (a) shows the annotation window listing gesture categories with a button to add new categories. Panel (b) displays saved instances of a "forget" gesture with an option to add new instances. Panel (c) shows the annotation interface with synchronized video playback and sensor reading visualization, featuring a timeline scrubber for segment selection. Panel (d) presents a visualization of all annotated gestures.}
    \caption{Screenshots of the app showing the gesture annotation process. (a) The ``Annotate Data'' window shows the available gesture categories. Users can create a new category by tapping on the ``+ NEW GESTURE CATEGORY'' button. A communicative message can be assigned to each of the categories. A tap on ``forget'' will open the window shown in (b), which lists the annotated instances of the ``forget'' gesture. New annotations of the ``forget'' gesture can be added by clicking on ``+ NEW GESTURE INSTANCE.'' (c) The annotation window with the selected video and corresponding sensor readings. Users can select a segment that represents the ``forget'' gesture in the timeline scrubber and save the new instance using the save button at the top-right in (c). (d) A visualization of annotated gestures in the app.}
    \label{fig:annotate_data}
\end{figure*}

\subsection{Initial Setup}
We use the Movesense IMU sensor\footnote{\url{https://www.movesense.com/movesense-hr-datasheet/}}\textsuperscript{,}\footnote{\url{https://www.movesense.com/product/movesense-sensor-hr/}}, a lightweight (9.4g) device with Bluetooth Low Energy (BLE). The sensor provides 9-DoF movement data (accelerometer, gyroscope, magnetometer), but we rely only on accelerometer and gyroscope readings to avoid magnetic field noise.

The sensor pairs with any Bluetooth-enabled device. For our study, we used Android phones (e.g., Pixel 8a). As shown in Fig.~\ref{fig:initial_setup}, users or aides can pair sensors from the AllyAAC app’s home screen (a), scan for nearby devices (b), and begin recording after securing the sensor on the wrist (c). These design choices reflect the input of the advisors about the need for flexible mounting options (wrist- or wheelchair-based). For now, the app functions as a proxy for aided AAC. In practice, its features will be integrated into mainstream AAC devices.

\subsection{Recording Template Gestures}
To support recognition of personalized communicative gestures (e.g., ``forget it''), users recorded gesture templates while wearing the sensor. The ``Record Data'' window (Fig.~\ref{fig:initial_setup}c) captured synchronized video and IMU data at 50 Hz, fast enough for gesture recognition while avoiding lag or battery strain.

Each gesture was recorded at least three times to capture intra-gesture variation. At minimum, users recorded one ``idle'' gesture and one or more communicative gestures (e.g., ``I’m hungry,'' ``go faster,'' ``hello''). The video served as a reference for later annotation. 
We additionally collected longer recordings in which users performed multiple gestures within the same session to capture natural transitions between gestures.

\begin{figure*}[t!]
    \centering
    \includegraphics[width=0.95\textwidth]{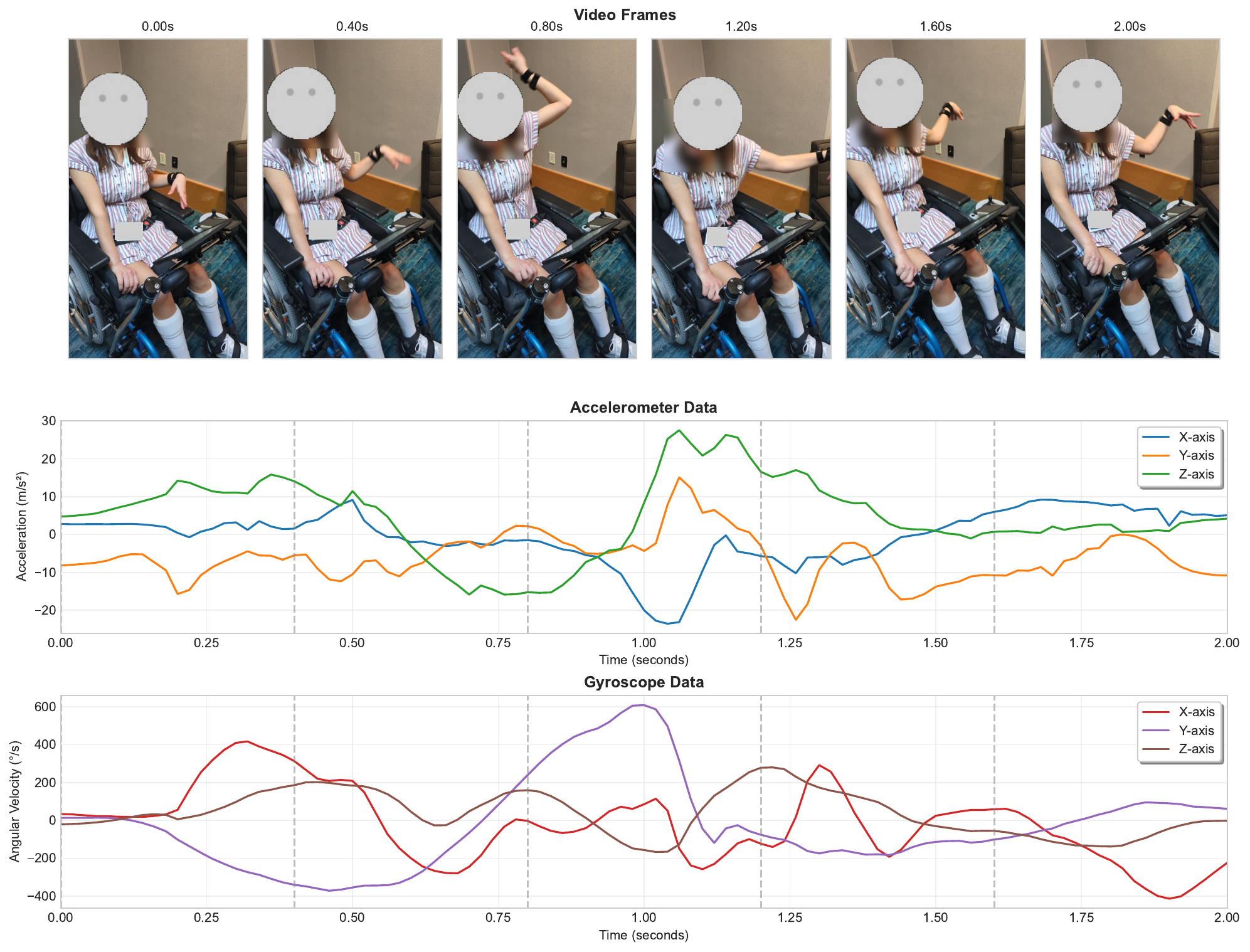}
    \Description{A visualization showing synchronized video frames and sensor data for a "forget it" gesture performed by participant P03. The top row displays five video frames with timestamps. The bottom two rows show corresponding accelerometer and gyroscope readings plotted as line graphs with three axes each, aligned temporally with the video frames.}
    \caption{Synchronized video frames with accelerometer and gyroscope data showing the movement and inertial measurements of the ``forget it'' gesture of P03. The top row shows the video frames with accurate timestamps displayed above each frame. The bottom two rows show the accelerometer and gyroscope sensor readings synchronized with the video timestamps.}
    \label{fig:gesture_visualization}
\end{figure*}

% \rrh{Semi-Automatic} 
\subsection{Gesture Annotation}
\label{subsec:gesture_annotation}
Users or aides annotated gesture instances within the app (Fig.~\ref{fig:annotate_data}). The annotation interface allowed browsing of recorded videos, scrubbing timelines, and marking gesture start and end points while viewing synchronized sensor traces.
Users found it helpful to inspect sensor patterns alongside videos, such as gyroscope variance or accelerometer directionality. For example, during the ``forget it'' gesture (Fig.~\ref{fig:gesture_visualization}), the gyroscope's Y-axis showed strong rotational movement, while the accelerometer Z-axis remained near $-9.8$ m/s² due to gravity, consistent with a downward wrist orientation.

However, annotating every instance of all gesture categories becomes time-consuming and repetitive as the number of categories or recorded samples increases. To reduce this burden, we designed a semi-automatic annotation pipeline that requires users to provide only coarse annotations, which the pipeline then refines automatically.
The pipeline expects a gesture video recording that contains multiple instances of the same gesture category (A) separated by an idle state (B):
\aptLtoX[graphic=no,type=html]{
\begin{itemize}
\item [\textbf{A}:] A communicative gesture with clear hand movement
\item [\textbf{B}:] An idle state representing the default hand position
\end{itemize}
}{
\begin{itemize}
\item [\textbf{A}:] A communicative gesture with clear hand movement
\item [\textbf{B}:] An idle state representing the default hand position
\end{itemize}
}

Users coarsely annotate a region of video where A and B occur repeatedly. From this coarse annotation, the pipeline segments each occurrence of A and B within the region using MediaPipe~\cite{zhang2020mediapipe}.

% \subsectionrr{\hst{Model Creation}}
% \rx{People perform the same gesture in different ways, shaped by motor ability, posture, and habit. As a result, gesture recognition requires models that are personalized to each user. We introduced two approaches: a rule-based, non–deep learning model that can be set up in under 10 minutes but offers limited performance, and a deep learning model that achieves higher accuracy but requires more compute to train for each user.}

%%%%%%%%%%%%%%%%%%%%%% LCSS Eq from old version
% To convert continuous sensor readings into discrete sequences for LCSS, we apply $k$-means clustering with 20 clusters ($k=20$), following the recommendation from \cite{nguyen2012improving}. 
% Each datapoint is assigned a symbolic label $C_1 \ldots C_{20}$, producing sequences like $C_3\, C_4\, C_1\, \ldots\, C_{12}$. Fig.~\ref{fig:k_means} shows this clustering process. These symbolic sequences form the templates for future comparison.
% \paragraph{Similarity Scoring}
% Let $S = (s_1, \ldots, s_m)$ be the symbol sequence from a new window of sensor data, and let $T = \{T^1, \ldots, T^n\}$ be $n$ gesture templates. For each $T^j$, we compute:
% \[
% \sigma_j = \frac{\text{LCSS}(S, T^j)}{\max(|S|, |T^j|)}
% \]
% where $\text{LCSS}(S, T^j)$ is the length of the longest common subsequence. The recognized gesture is the one with the highest $\sigma_j$, provided it exceeds a threshold $\tau = 0.4$. Otherwise, the system defaults to the idle state.
%%%%%%%%%%%%%%%%%%%%%% end LCSS Eq from old version

\subsection{Baseline Model: Rule-Based}
\label{subsec:baseline_model}
As the baseline, we used a rule-based model~\cite{wilkinson2025consideration} that uses the Longest Common Subsequence (LCSS) similarity metric~\cite{nguyen2012improving}. IMU signals were discretized into clusters using K-means~\cite{mcqueen1967some} and compared against pre-recorded gesture templates. This is a multi-class classifier that outputs one of $M+1$ classes, where $M$ is the number of gesture categories and the additional class represents the idle state. Model creation with this method was fast, typically in less than 10 minutes, and was performed by users (or their aides), making it practical for rapid personalization. However, performance degraded as the number of gesture categories increased, reaching only about 60\% $F_1$-score in our lab tests.  

\begin{figure*}[t!]
    \centering
    \includegraphics[width=0.99\textwidth]{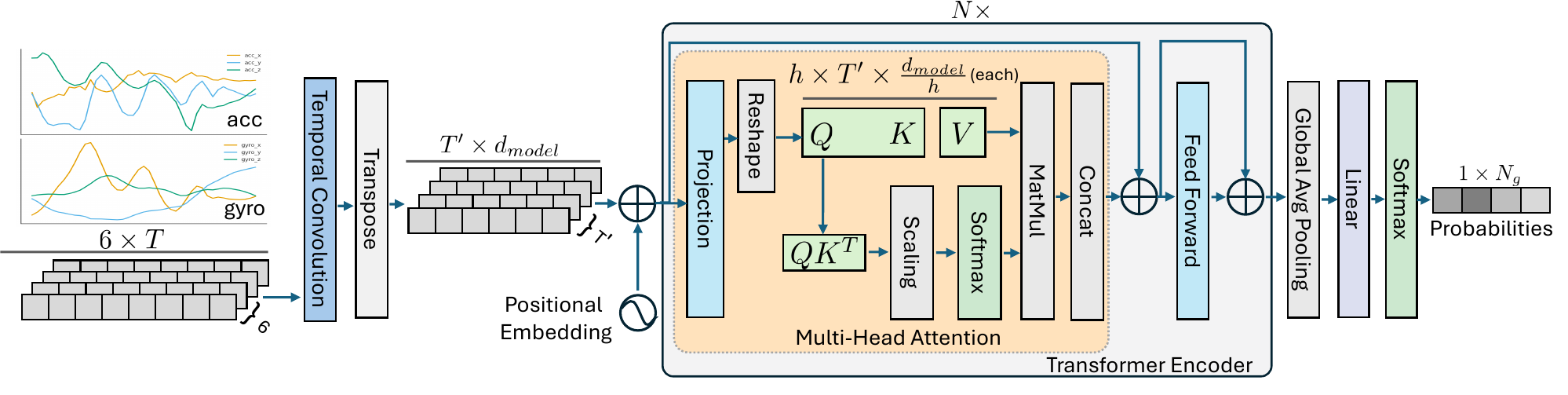}
    \Description{A diagram illustrating the large model architecture for gesture recognition. The pipeline flows left to right: 6-channel IMU input undergoes temporal convolution, then projection to tokens, followed by addition of positional embeddings. The sequence passes through transformer encoder blocks, then global average pooling, and finally a linear layer with softmax to produce gesture class probabilities.}
    \caption{Architecture used for our large gesture recognition model. The model takes a time series of 6-channel IMU signals (accelerometer and gyroscope), applies temporal convolution to extract local motion features, and projects the result into a token sequence. Positional embeddings are added before passing the sequence through the transformer encoder blocks to model temporal dependencies. The transformer outputs are pooled across time using global average pooling, and a final linear layer and softmax map the result to gesture class probabilities.}
    \label{fig:model_arch}
\end{figure*}

\begin{figure*}[t!]
    \centering
    \includegraphics[width=0.9\textwidth]{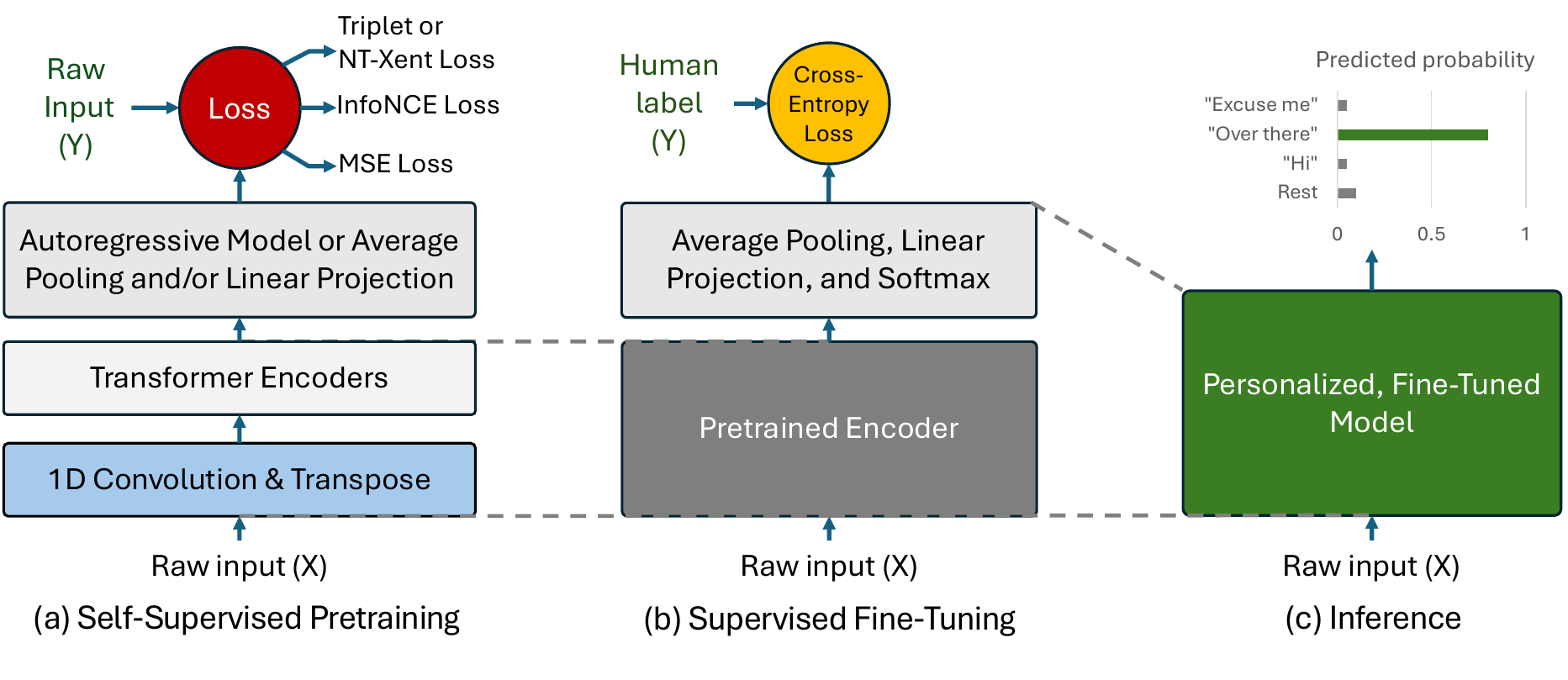}
    \Description{A three-part diagram showing the model training pipeline. Panel (a) shows the model architectures used in self-supervised pretraining with three loss functions: Triplet/NT-Xent, InfoNCE, and MSE. Panel (b) shows the supervised fine-tuning stage using cross-entropy loss. Panel (c) illustrates real-time inference where the trained model outputs predicted probabilities for gesture categories.}
    \caption{
    Illustration of the complete model training pipeline. (a) Simplified model architecture for self-supervised pretraining. For three different tasks, three different loss functions (Triplet~\cite{schroff2015facenet} or NT-Xent~\cite{chen2020simple}, InfoNCE~\cite{oord2018representation}, and MSE) are used. (b) Simplified architecture of the model used in supervised fine-tuning. In this stage, we use cross-entropy loss. (c) Finally, the fine-tuned model is used for real-time inference, where the model outputs predicted probabilities for each gesture category.
    }
    \label{fig:training_pipeline}
\end{figure*}

\subsection{Large Model: Deep Learning Based}
\label{subsec:large_model}
Similar to the baseline, our large model is a multi-class classifier with $M+1$ output classes. It is based on deep learning (uses a Transformer encoder~\cite{vaswani2017attention}) and processes sequential IMU data (accelerometer and gyroscope) to output the probabilities of gesture categories (Fig.~\ref{fig:model_arch}). The input is a time series of shape $6 \times T$ ($T=120$), where each timestep contains six sensor channels: $\text{acc}_x$, $\text{acc}_y$, $\text{acc}_z$, $\text{gyro}_x$, $\text{gyro}_y$, and $\text{gyro}_z$.

The sequence is first normalized, then passed through a one-dimensional temporal convolution (kernel size $k=8$ and stride $s=1$) layer that extracts local motion features and projects them into a feature space of size $d_{\text{model}}$ (set to 128). The convolution output is transposed to form token embeddings of shape $T' \times d_{\text{model}}$, where $T'$ is the temporal length after convolution. Positional embeddings are added to preserve temporal order.  

The embeddings are passed through $N$ (set to 3) transformer encoder blocks~\cite{vaswani2017attention}. Each block contains multi-head self-attention, where the attention weights are computed from query ($Q$) and key ($K$) vectors and applied to value ($V$) vectors, producing outputs for each of the $h$ (set to 4) heads as shown in Fig.~\ref{fig:model_arch}. These are aggregated and passed through a feed-forward network with skip connections. Finally, the output of the transformer blocks is pooled across time with global average pooling and passed through a fully connected (linear) layer that maps the representation to gesture class logits, which are then converted to probabilities of size $1 \times N_g$ using softmax. Here, $N_g=M+1$ is the number of gesture categories (including idle) for that user.

\begin{table*}[t!]
\centering
\begin{tabular}{l|l|p{12.6cm}}
\hline
\textbf{ID} & \textbf{Gestures} & \textbf{Intention} \\
\hline
\multirow{4}{*}{P01} & Drink & The user moves their fist toward their face, mimicking the action of drinking. \\
\cline{2-3}
 & Go fast & The user moves their hand up and down on their side, signaling a desire for increased speed. \\
\cline{2-3}
 & Over there & The user extends their hand outward, moving it from inside to out, indicating a specific direction. \\
\cline{2-3}
 & Yes & The user raises their hand up and then brings it down to indicate affirmation. \\
\hline
\multirow{2}{*}{P02} & Hungry & The user moves their fist toward their mouth, signaling a need for food. \\
\cline{2-3}
 & Point & The user extends their finger to indicate a specific object or direction. \\
\hline
\multirow{3}{*}{P03} & Hungry & The user moves their hand toward their mouth and then brings it back, signaling a desire for food. \\
\cline{2-3}
 & Happy & The user lifts their hand up and moves it back, expressing joy or positivity. \\
\cline{2-3}
 & Forget & The user raises their hand and then flicks it away, signaling dismissal or letting go of something. \\
\hline
\multirow{3}{*}{P04} & Hello & The user raises their hand and waves, signaling a greeting. \\
\cline{2-3}
 & Look at that & The user extends their hand outward briefly and then brings it back, directing attention to something. \\
\cline{2-3}
 & Need to use the bathroom & The user waves their hand and then taps their lower abdomen, signaling the need to use the restroom. \\
\hline
\multirow{4}{*}{P05} & Letter B & The user traces the letter "B" in the air using their hand. \\
\cline{2-3}
 & Letter P & The user traces the letter "P" in the air using their hand. \\
\cline{2-3}
 & Letter R & The user traces the letter "R" in the air using their hand. \\
\cline{2-3}
 & Letter T & The user traces the letter "T" in the air using their hand. \\
\hline
\multirow{3}{*}{P06} & Need to use the bathroom & The user points a finger toward their lower body, signaling the need to use the restroom. \\
\cline{2-3}
 & Drink & The user positions their hand as if holding a cup and moves it slightly upward, signaling the action of drinking. \\
\cline{2-3}
 & Get out of the way & The user flicks their hands outward, signaling for someone to move aside. \\
\hline
\multirow{4}{*}{P07} & Don't Like & The user covers their ears with their hands, signaling they don't like it. \\
\cline{2-3}
 & Happy & The user rapidly waves their hands in a clap-like motion, expressing excitement or happiness. \\
\cline{2-3}
 & I want & The user extends their arm forward to indicate they want something. \\
\cline{2-3}
 & Love it & The user brings their hand to their mouth in a delighted motion, expressing that they love it. \\
\hline
\multirow{3}{*}{P08} & Excuse me & The user holds a fist in front of their face and quickly swings it rightward, with the elbow moving to the right, to signal ``excuse me.'' \\
\cline{2-3}
 & One moment & The user raises their index finger upward in front of their body to signal ``one moment.'' \\
\cline{2-3}
 & I have a question & The user raises their hand to signal they have a question. \\
\hline
\end{tabular}
\Description{A table listing each participant's (P01–P08) personalized gesture types and their associated communicative intentions or messages.}
\caption{Participants' gesture types and their intentions.}
\label{tab:gestures}
\end{table*}

% To account for the scarcity of annotated data, we train this model in two stages: (i) self-supervised pretraining (see Fig.~\ref{fig:training_pipeline}a) using objectives such as contrastive learning~\cite{tang2020exploring,chen2020simple}, masked reconstruction~\cite{haresamudram2020masked}, or contrastive predictive coding (CPC)~\cite{haresamudram2021contrastive}, which involves predicting representation of future timesteps; and (ii) supervised fine-tuning (see Fig.~\ref{fig:training_pipeline}b)  for gesture classification using limited annotated gestures (3–20 examples per category). 
To account for the scarcity of annotated data, we train this model in two stages: 
(i) self-supervised pretraining (see Fig.~\ref{fig:training_pipeline}a) using objectives such as contrastive learning~\cite{tang2020exploring,chen2020simple}, masked reconstruction~\cite{haresamudram2020masked}, or contrastive predictive coding (CPC)~\cite{haresamudram2021contrastive}, where CPC involves predicting representations of future timesteps; and 
(ii) supervised fine-tuning (see Fig.~\ref{fig:training_pipeline}b) for gesture classification using limited annotated gestures (3–20 examples per category).
Fig.~\ref{fig:model_arch} illustrates the model architecture used in the second stage (gesture classification). For the first stage, different strategy-specific heads and loss functions replace the classification head (pooling, linear projection, and softmax) as shown in Fig.~\ref{fig:training_pipeline}. 
% Fig.~\ref{fig:training_pipeline} shows all the steps of our training pipeline. 
Once the model was fully trained, it was used for inference (see Fig.~\ref{fig:training_pipeline}c). Next, we briefly discuss the three self-supervised training strategies.

\subsubsection{Contrastive Learning} 
\label{subsubsec:contrastive}
% Contrastive learning minimizes the distance between the same gesture pairs (one gesture instance and its augmented versions) and increases the distance between that gesture and other gestures.
Contrastive learning minimizes the distance between representations of positive pairs (a gesture instance and its augmentations) and maximizes the distance between negative pairs corresponding to different gesture instances.
The transformer encoder processes each gesture (original instance, its augmentations, and others); its output is pooled and projected to create feature vectors for anchor, positive, and negative samples.
The model is optimized using Triplet~\cite{schroff2015facenet} or NT-Xent~\cite{chen2020simple} loss.

% The transformer encoder's output is pooled and linearly projected to create feature vectors for the original instance, positive (augmentations), and negative (other instances) samples.
% Triplet loss~\cite{schroff2015facenet}. 

\subsubsection{Contrastive Predictive Coding (CPC)} 
\label{subsubsec:cpc}
Inspired by CPC~\cite{haresamudram2021contrastive}, we train the model to predict representations of the next $L$ timesteps from a randomly sampled timestep within each input sequence (we use $L=16$ for our experiments). The model is optimized using the InfoNCE loss~\cite{oord2018representation}.

% Inspired by CPC~\cite{haresamudram2021contrastive}, we train the model to predict representations of the next $L$ timesteps from a randomly sampled timestep within each input sequence (we use $L=16$ for our experiments). The model is optimized using the InfoNCE loss~\cite{oord2018representation}.

% To optimize the model, InfoNCE~\cite{oord2018representation} loss is used.
% Inspired by Contrastive Predictive Coding~\cite{haresamudram2021contrastive}, we train the model to take input sequences and predict representations of the next $L$ future timesteps within the sequence (where $L=16$ is the prediction horizon). To optimize the model, it uses InfoNCE~\cite{oord2018representation} loss.

\subsubsection{Masked Reconstruction} 
\label{subsubsec:masked_recon}
15\% of the timesteps in the input are randomly masked. This masked input is passed to the encoder of the large model. The encoder's output is then projected back to the input dimension, and the MSE loss measures the deviation between the model's reconstructed output and the original unmasked values at the masked positions. This approach optimizes the model to reconstruct the masked data points from the surrounding context.

\subsection{Model Training and Deployment}
\label{subsec:model_deploy}

The rule-based baseline model does not require training; users can configure it directly on their device in under 10 minutes once the data preparation (recording, annotation, and communicative message assignment) is done.  

For the deep learning model, we implemented the architecture in PyTorch and trained it on CUDA-enabled machines. Data collected on the mobile device was exported using the ``Export Data'' option (Fig.~\ref{fig:initial_setup}a), which packaged both videos and sensor logs as a ZIP file. This file was transferred to the training machine, where end-to-end training required approximately 2–3 hours, depending on the dataset size. Training included both self-supervised pretraining and supervised fine-tuning. We trained for 200 epochs in self-supervised pretraining and 25 epochs in fine-tuning.  

\begin{figure*}[t!]
    \centering
    \includegraphics[width=0.99\textwidth]{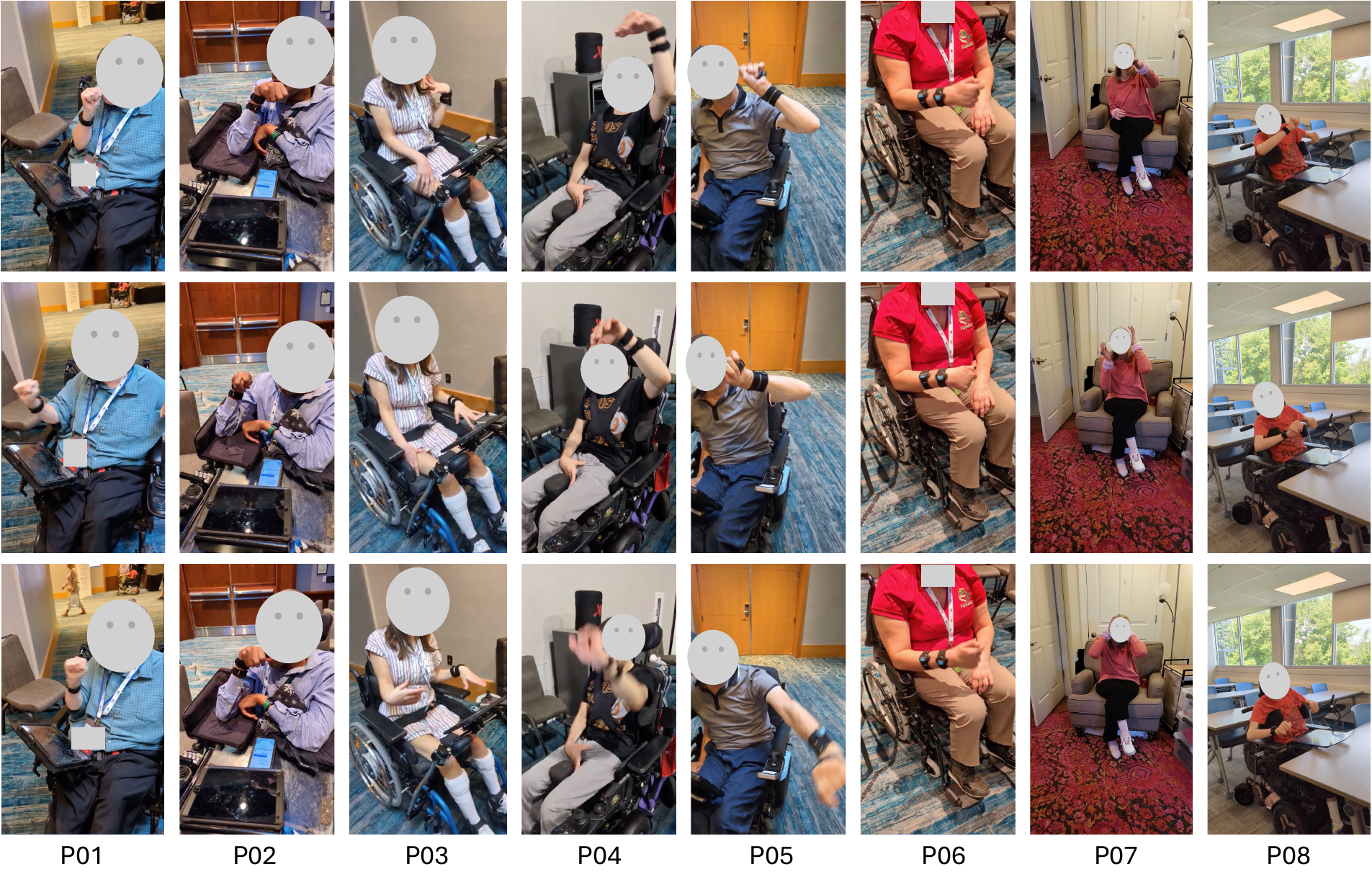}
    \Description{A grid of video frames showing gestures from eight motor-impaired participants (P01–P08). Each column contains three frames capturing one participant performing their first gesture from Table 1. The gestures range from P01's "Drink" gesture on the left to P08's "Excuse me" gesture on the right.}
    \caption{Video frames of each motor-impaired participant’s (P01–P08) first gesture as reported in Table~\ref{tab:gestures}. Each column corresponds to one participant, showing three frames that capture the execution of their first gesture. For example, the first column shows P01's ``Drink'' gesture and the last column shows P08's ``Excuse me'' gesture.}
    \label{fig:motor_impaired_participants_gesture}
\end{figure*}

After training, we converted the model to the \texttt{tflite} format for deployment on mobile devices using \textit{Google AI Edge Torch}\footnote{\url{https://github.com/google-ai-edge/ai-edge-torch}}. 
The final model had 0.6M parameters and ran in real-time on Android phones after the conversion.
The AllyAAC app supports importing a \texttt{tflite} model and its metadata as a ZIP file via the ``Import Model'' option (Fig.~\ref{fig:initial_setup}a). Metadata specifies the model’s input/output shapes, gesture classes, and normalization factors (mean and standard deviation) used during training. Once imported, the model runs directly within the ``Run Model'' window (Fig.~\ref{fig:initial_setup}d), enabling real-time gesture recognition. A sliding window of $T=120$ timestamps ($\approx$2.4 s) with 50\% overlap was applied to continuous streams. As with aided AAC systems, the app announces the message associated with each detected gesture using a built-in TTS engine. 
The app also implements the clutching mechanism (requested during participatory design) through a toggle button (Fig.~\ref{fig:initial_setup}f), which allows users to pause and resume gesture recognition on demand.

% \section{Qualitative Evaluation of AllyAAC  \hst{in the Wild} in Natural Settings}
% \label{sec:qual_eval}
% By the end of the 18-month design process, AllyAAC had reached a stable state. Since many use cases involved crowds and unfamiliar partners, the research team met the community advisors in person at a conference on assistive technology. This setting allowed us both to assess how advisors benefited from AllyAAC in practice and to evaluate the system with a wider group of AAC users through a field study in \rx{the wild} natural settings.  
\section{Qualitative Evaluation of AllyAAC in Social Context}
\label{sec:qual_eval}
By the end of the 18-month design process, AllyAAC had reached a stable state. Since many use cases involved crowds and unfamiliar partners, the research team met the community advisors in person at a conference on assistive technology (ATIA'25). This setting allowed us both to assess how advisors benefited from AllyAAC in practice and to evaluate the system with a wider group of AAC users through a field study in natural settings. 

\subsection{Participants}
In total, fourteen adults participated in our field study.
Eight had lower-body motor impairments of varying degrees and relied on wheelchairs full-time (P01--P08). Two of these eight participants were community advisors (P07 and P08) who wanted to interact with others at the conference. Six of the eight motor-impaired participants (P01--P05 and P07) were non-verbal and relied exclusively on AAC for communication. The other two can speak and use AAC devices infrequently.
Two participants reported visual impairments: P06 was blind, and P07 had cortical visual impairment. To accommodate visually impaired users, our mobile application was designed to be accessible with screen readers.
We also use a darker background to accommodate users with cortical visual impairment who are sensitive to bright light.
While the gesture annotation process required visual feedback and assistance from a sighted aide, once the personalized model was trained, the system became fully accessible for independent use.
The remaining six participants were non-disabled individuals—aides, care providers, or researchers who frequently interact with AAC users (P09–P14). 
All participants were over 18 years old. Table~\ref{tab:gestures} lists P01–P08’s gesture sets and associated intentions, and Fig.~\ref{fig:motor_impaired_participants_gesture} shows images of each participant’s first gesture.  

The inclusion of community advisors, who had deep familiarity with the system, brought both strengths and limitations. On the one hand, they served as expert users whose fluent interaction provided a useful point of comparison with first-time users. Their consistency in performing gestures resulted in cleaner and less noisy data. On the other hand, their expertise may have biased measures tied to system familiarity or learning curve. To avoid inflating performance, we did not report time-based efficiency metrics (e.g., task completion time, first-gesture latency). Instead, our evaluation emphasized recognition precision, $F_1$-scores, and qualitative feedback, which are more robust to differences in user experience.  

The inclusion of non-disabled participants offered another perspective. As aides, caregivers, or researchers, they often mediate AAC interactions. Their feedback provided insight into how AllyAAC might be introduced and supported in real-world contexts, particularly for new AAC users.  

\subsection{Materials}
We used an Android Pixel 8a phone as the primary device (Section~\ref{sec:workflow}), paired with a wrist-worn IMU sensor that transmitted motion data via Bluetooth. A second IMU sensor was worn on the same wrist to collect unstructured data. This was paired with a secondary Android Pixel 7, which also video-recorded the full session. Each data collection session lasted 45–60 minutes.

\subsection{Procedures}
\label{subsec:study_procedure}
After providing consent, each session began with participants describing communicative gestures they currently use or would like to use in daily interactions. Participants were encouraged to suggest gestures that carried personal meaning, were already familiar to them, or served functional purposes in home or community settings. There was no set limit on the number of gestures: some participants explored two or three, while others identified more depending on interest, comfort, and time.  

Once a gesture was selected, the researcher placed the IMU sensor on the participant’s preferred wrist and recorded at least three repetitions of the gesture using the app. This was followed by recordings of “rest” trials to capture the participant’s default posture without intentional gesture input.  
The app simultaneously recorded video and IMU sensor data at 50 Hz, with both streams automatically timestamped to ensure precise synchronization. In practice, each video contained multiple instances of only one communicative gesture with idle/rest periods in between to separate individual instances. Videos were recorded such that the hand used for making the gesture was always visible in the scene (Fig.~\ref{fig:motor_impaired_participants_gesture} shows example frames from recorded videos). This structured video recording made the annotation process easier.

After each gesture was recorded, the researcher collaborated with the participant to annotate the corresponding video and sensor segments, ensuring that the data accurately reflected the intended movement. The participant then assigned a spoken message for the app to output when that gesture was recognized (e.g., ``I’m hungry,'' ``Look over there,'' or ``Please move''). This message is customizable and can be easily assigned and modified from the app. During the study, recording gestures, annotating them, and assigning communicative messages to each gesture category took less than 8 minutes. On average, each participant had 3--4 gestures. Therefore, the entire process, including data preparation and model building, took 30 minutes on average per participant.

Once a gesture was annotated and mapped to a message, participants tested the prototype by performing the gesture to see whether it was correctly recognized and spoken using the baseline rule-based model. This process was repeated for each gesture the participant wished to explore. 
Participants were encouraged to test gestures multiple times and provide real-time feedback on recognition accuracy, ease of performance, and perceived utility. At the end of each session, they also shared reflections on how the system might be used in real-world settings and how the app or process could be improved. 
During this stage, only the baseline model was tested since it can be configured in under 10 minutes once the data annotation is done. Testing the large model was not feasible during the field study, as training personalized models for each individual required approximately 2--3 hours. We therefore conducted an offline evaluation of the transformer model by measuring $F_1$-scores and through human evaluation (see Section~\ref{sec:quant_eval}).

% Data Analysis. Throughout the session, the app continuously logged IMU sensor data and video recordings of gesture trials. A trial was marked as a success if the recognized gesture triggered the correct spoken message. A failure occurred when an incorrect message was spoken (false positive) or when the gesture went unrecognized (false negative).
% Once gestures were annotated and mapped to different communicative messages, the recorded data were transferred to a desktop for model training. Data transfer and model training were performed by the authors. Deploying a system that is under development on participants’ phones was not feasible, particularly models that had only been evaluated with quantitative metrics (e.g., $F_1$ score). Accordingly, we built a user interface for human-centered evaluation of the model. Six non-disabled participants took part in this assessment (P09, P10, P11, P12, P14, and one author who did not participate in data collection). Procedural details and results of this evaluation are provided in Section~\ref{subsec:human_eval}. This evaluation enabled us to assess model behavior with human judgment and iteratively improve the model before transferring it to participants’ phones for in-the-wild use.

\subsection{Data Analysis}
Since participants were encouraged to use gestures naturally rather than in rigid repetitions, the number of gestures and trials per gesture varied across sessions. In some cases, participants abandoned gestures that proved difficult to annotate or distinguish; in others, they prioritized evaluating usability over maximizing the number of trials. This flexible structure allowed us to assess feasibility under conditions that approximate real-world use.  
After cleaning, the dataset included 3 hours and 35 minutes of synchronized multimodal data (video + IMU), totaling more than 600,000 data points. We release this dataset to the AI and accessibility research communities~\footnote{Our dataset and code are available at: \url{https://github.com/Imran2205/AllyAAC}}.  
For open-ended feedback, we used an iterative coding process to identify recurring themes and refine categories across multiple passes. This approach enabled us to connect technical feasibility with participants’ lived experience on gesture usability and system integration.  

\begin{figure*}[t!]
    \centering
    \includegraphics[width=0.9\textwidth]{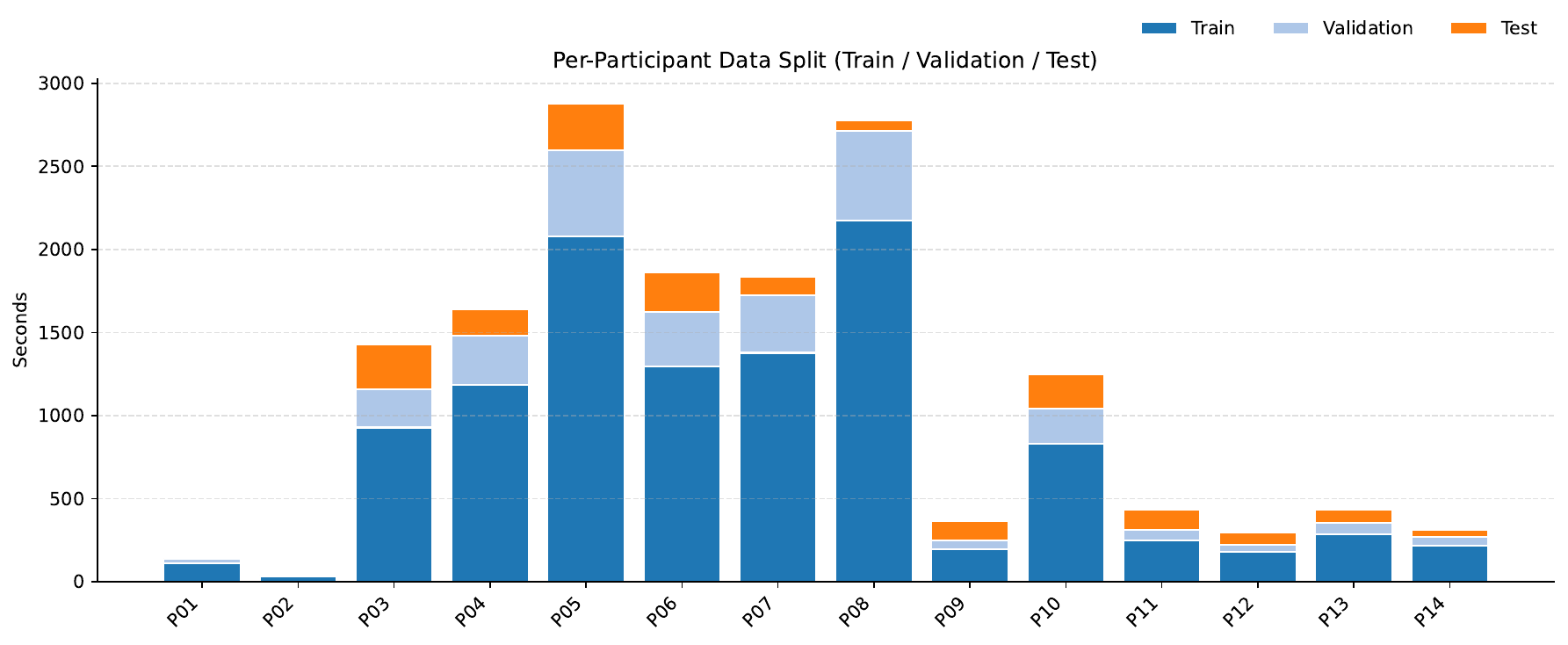}
    \Description{A vertical bar chart showing the duration in seconds of IMU data used for training, validation, and testing for each participant. Each bar is segmented into three colors representing the three data splits.}
    \caption{Duration (in seconds) of IMU data used for training, validation, and testing for each participant.}
    \label{fig:data_distribution}
\end{figure*}

\subsection{Observations and Findings}
Based on participants’ feedback and our observations, we identified recurring themes about what worked, where friction appeared, and how people envisioned using AllyAAC in everyday contexts.

\subsubsection{Gesture Selection and Recording}
Participants selected gestures that felt intuitive or were already part of their communication routines (see Table~\ref{tab:gestures}). Most recorded three to five gestures, though some attempted more. Common gestures included pointing, mimicking drinking, flicking, waving, and directional swipes.  

Although the recording interface was designed to be self-guided, most participants required assistance, particularly in transferring data for model training and importing the trained model for real-time use. Several aides expressed interest in using the tool themselves to support customization for their clients.

\subsubsection{Workflow Usability}
Participants responded positively to the overall pipeline, especially the annotation interface that displayed sensor traces alongside video. Sensor placement also influenced usability. While all participants wore the IMU on the wrist, some questioned whether the upper arm, back of the hand, or palm might be better. One participant asked for sensors on both arms, noting, \textit{``it’s easier for me to move one hand than the other depending on the chair position.''} Setup in general required assistance, particularly for sensor placement and navigation of small smartphone screens. One participant with limited dexterity suggested using a tablet to reduce touch precision demands. 
Participants found the clutching mechanism useful for preventing speech output during non-communicative hand movements.

\subsubsection{Participants’ Reflections}
Once participants experienced a working gesture-to-speech mapping, they generated ideas for real-world use cases. One proposed air-drawing a ``W'' to quickly indicate a bus destination. Another (P06) envisioned a flick gesture to say ``move out of the way'' while driving his power wheelchair.  

Others emphasized the broader value of translating movement into speech. P06 remarked, \textit{``Using this app would make it easier on everyone.''} P02 said, \textit{``It would be great, especially with unfamiliar communication partners when you don’t have access to your AAC device.''} Several expressed a desire for greater autonomy, including doing their own annotations, customizing speech messages, and using larger devices.  

Participants also pointed toward directions for future development: sequencing gestures to create multi-part messages, assigning meaning based on context or location (e.g., home vs. public), using repetition or speed to signal urgency, and expanding the gesture vocabulary beyond what researchers initially imagined.

\subsubsection{Annotation Empowered, Data Transfer Did Not}
Participants consistently described annotation as empowering. Seeing their own movements visualized gave them confidence and a sense of ownership. In contrast, data transfer and model import were confusing and burdensome. Exporting recordings and sending them to the research team, often via email or cloud storage, was time-consuming. Importing trained models back into the app required multiple manual steps. Participants agreed these processes should be automated in future systems, with synchronization and deployment handled seamlessly in the background.
% \subsubsection{Self-Supervised Pretraining Reduces Data Burden}
% While recording a few gestures was feasible, several participants lost interest or experienced fatigue when asked to record five or more. Building reliable recognition models requires a substantial amount of training data—including negative examples. Pre-training general-purpose gesture embeddings using self-supervised learning (e.g., contrastive learning~\cite{tang2020exploring}, masked reconstruction~\cite{haresamudram2020masked}, or CPC~\cite{haresamudram2021contrastive}) helped to achieve reasonable performance with limited data annotation.

\begin{figure*}[t!]
    \centering
    \includegraphics[width=0.95\textwidth]{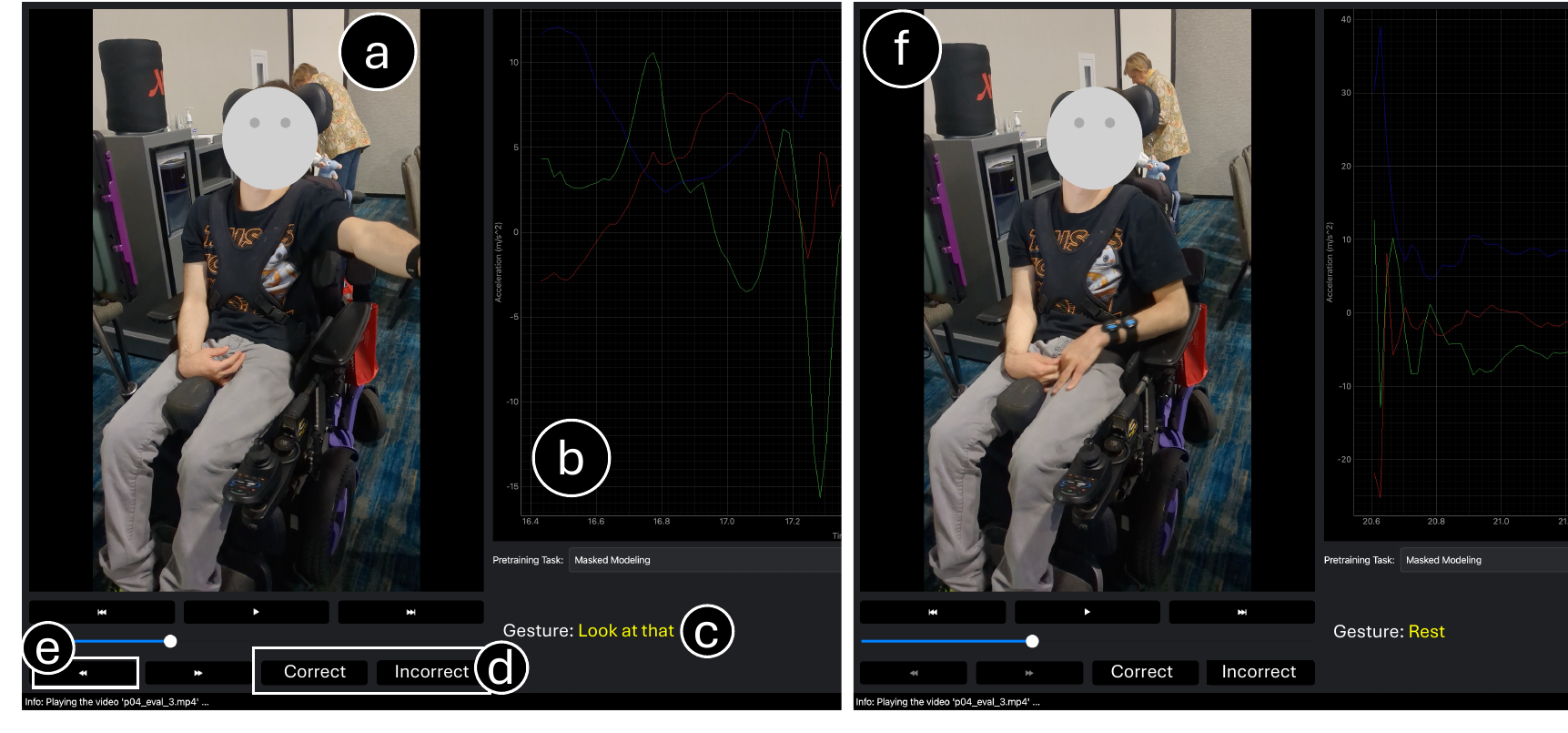}
    \Description{A screenshot of the human evaluation tool interface. Panel (a) shows a video player displaying a participant performing a gesture. Panel (b) displays a sensor stream visualization of the last two seconds of data. Panel (c) shows the model's predicted gesture label. Panel (d) contains buttons for marking predictions as correct or incorrect. Panel (e) shows a replay button. Panel (f) displays an example frame with the participant's hand at rest.}
    \caption{Tool developed for conducting the human evaluation. (a) Video player displaying a participant performing a personalized gesture. (b) Sensor stream visualization showing the last two seconds of data aligned with the current video timestamp. (c) Model-predicted gesture for the current sliding window (“Look at that” in this example). (d) Buttons for marking the prediction as correct or incorrect. (e) Replay button, which replays the current gesture segment. (f) Example frame where the participant’s hand is in the rest position.}
    \label{fig:eval_ui}
\end{figure*}

\section{Quantitative Evaluation of AllyAAC}
\label{sec:quant_eval}
By the end of the study, recorded data were used to prepare and evaluate both the rule-based baseline and the deep-learning-based model (large model). We quantitatively evaluated AllyAAC in five different aspects: 1) the effectiveness of semi-automatic annotation; 2) the effectiveness of different pretraining strategies; 3) the introspection of large models; 4) the classification performance of the baseline and the large model; and 5) human ratings of the best large models for an individual.

\subsection{Procedures}
To carry out the evaluation, we first annotated the entire dataset using the semi-automatic annotation pipeline.
The semi-automatic annotation pipeline benefited from the structured recording format discussed in Section~\ref{subsec:study_procedure}, where each video contained only one communicative gesture separated by idle/rest gestures, and automatically annotated gestures with minimal human intervention (required only for verification). The few videos that deviated from this format were handled through coarse manual annotation to mark video segments that met the criteria for automatic annotation (see Section~\ref{subsec:gesture_annotation}), followed by automatic annotation. This process produced 3--20 gesture instances per gesture category for each participant.

\subsubsection{Effectiveness of Semi-Automatic Annotation}
To evaluate the effectiveness of the semi-automatic annotation pipeline, we compared the annotation time and timestamps of automatically annotated gestures with those of human-annotated gestures. For this experiment, we selected 7 participants: 4 AAC users with motor impairments (P03, P04, P06, P08) and 3 able-bodied aides or researchers (P10, P11, P12). Two researchers manually annotated all gesture instances for these participants. We focused only on communicative gestures and excluded the rest gestures. In total, the researchers manually annotated 230 gesture instances across all participants, which were considered the ground truth.

\paragraph{Evaluation Metrics.} We computed the $F_1$-score to measure the reliability of automatic annotation, the mean IoU to quantify the temporal overlap between automatic and manual annotations, and the time saved by the semi-automatic annotation process (see Section~\ref{subsec:eval_annotation_pipeline}).

\subsubsection{Effectiveness of Pretraining Strategies}
To account for the limited annotated data, we pretrained the large model using three different strategies for each participant as discussed in Section~\ref{subsec:large_model}: contrastive learning, CPC, and masked reconstruction. The pretrained models were then fine-tuned with a handful of annotated data. This step produced three personalized gesture recognition models per participant.

\paragraph{Dataset Preparation for Large Models}
Our dataset comprised synchronized IMU and video recordings totaling 12,919 seconds.
% We break down the entire dataset into training, testing, and validation sets.
For each participant, we split recordings into training, validation, and test sets (as shown in Fig.~\ref{fig:data_distribution}). 
The training set was used for model training, the validation set for quantitative evaluation, and the test set for human evaluation.
To simulate realistic conditions, we used limited annotated data for supervised training: 3--20 samples per category for communicative gestures, while for rest gestures, the entire recordings of rest gestures were used with a sliding window approach, which did not require any annotation. The entire training set (without annotations) supported pretraining.
We lost a significant amount of data from P01 and P02 due to a technical error (sensor malfunction). 
The remaining data was sufficient to create the training and validation splits, with 3 instances per gesture category in the training set, which allowed us to train the models for them and quantitatively evaluate the models. However, the remaining data was not sufficient for human evaluation.
In total, the dataset includes more than 600,000 IMU data points. 
% We will release this dataset to the AI and accessibility research communities.

\paragraph{Evaluation Metrics.} We computed the $F_1$-scores of all three models for each participant to compare their scores and identify which strategy worked best for a specific type of motor pattern (Section~\ref{subsec:finding_pretraining}).

\begin{figure*}[t!]
    \centering
    \includegraphics[width=0.99\textwidth]{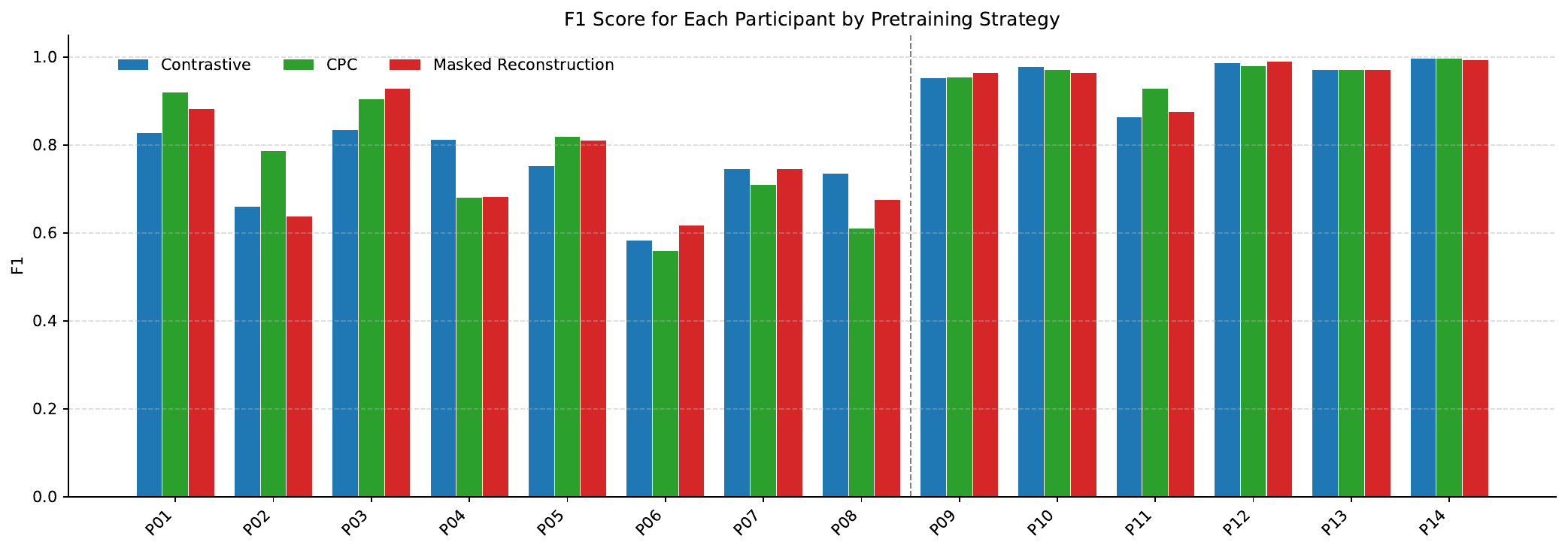}
    \Description{A grouped bar chart showing $F_1$-scores for large models across participants using three pretraining strategies (contrastive, CPC, and masked reconstruction). A vertical dashed gray line separates motor-impaired participants on the left from non-disabled participants on the right. Each participant has three bars representing the three strategies.}
    \caption{$F_1$-scores of large models for each participant across three pretraining strategies (contrastive, CPC, and masked reconstruction). The bars on the left of the dashed gray line represent scores for users with motor impairments, while the bars on the right represent scores for users without any motor impairment.}
    \label{fig:model_performance}
\end{figure*}

\subsubsection{Introspection of Large Models}
We analyzed the $F_1$-scores of the large model for motor-impaired users and able-bodied users to identify the gap in model performance between the two participant groups. To understand why this gap occurred, we visualized the learned embeddings from the transformer encoder of the best model for two motor-impaired participants (P03 and P06) and two able-bodied participants (P10 and P11) using t-SNE visualization~\cite{maaten2008visualizing} (see Section~\ref{subsec:introspection}).
% These scatter plots reveal the separability of gesture categories for each participant, allowing us to reason about the performance gap.

\subsubsection{Classification Performance}
% assess the ability of AllyAAC to recognize personalized gestures and 
We compared the $F_1$-scores of two models: (1) the baseline model, and (2) the large model. 
For the comparison, we selected the best large model for each participant among the three strategies based on their $F_1$-scores.
As described in Section~\ref{subsec:baseline_model} and ~\ref{subsec:model_deploy}, the baseline model offers rapid setup (under 10 minutes) but limited performance, while the transformer model requires longer training but provides higher $F_1$-scores. Evaluating both allows us to understand the trade-offs between speed of deployment and recognition performance.

\paragraph{Evaluation Metrics.} We computed the $F_1$-scores of both models to compare how well they recognize gestures and identify success and failure cases (see Section~\ref{subsec:finding_classification}).

\begin{table}[!ht]
\centering
\small
\setlength{\tabcolsep}{5pt}
\begin{tabular}{lcc}
\toprule
Participant & Baseline (rule-based) & Large Model \\
\midrule
P01 & 0.5293 & \textbf{0.9210} \\
\hline
P02 & 0.5556 & \textbf{0.7870} \\
\hline
P03 & 0.5349 & \textbf{0.9290} \\
\hline
P04 & 0.6211 & \textbf{0.8120} \\
\hline
P05 & 0.2154 & \textbf{0.8200} \\
\hline
P06 & 0.4310 & \textbf{0.6170} \\
\hline
P07 & 0.2639 & \textbf{0.7460} \\
\hline
P08 & 0.4333 & \textbf{0.7360} \\
\hline
P09 & 0.9153 & \textbf{0.9640} \\
\hline
P10 & 0.1966 & \textbf{0.9790} \\
\hline
P11 & 0.6538 & \textbf{0.9290} \\
\hline
P12 & \textbf{1.0000} & 0.9900 \\
\hline
P13 & 0.9487 & \textbf{0.9720} \\
\hline
P14 & 0.9407 & \textbf{0.9970} \\
\midrule
Mean & 0.5885 & \textbf{0.8714} \\
\bottomrule
\end{tabular}
\Description{A comparison table showing $F_1$-scores for each participant, comparing the baseline LCSS-based recognizer against the best large model (selected from contrastive, CPC, and masked reconstruction pretraining strategies). Bold values indicate the better-performing model for each participant.}
\caption{$F_1$-scores of models for each participant, comparing baseline (LCSS-based recognizer) and large model (best of each participant among pretraining strategies: contrastive, CPC, and masked reconstruction). Bold values indicate the better-performing model for each participant.}
\label{tab:f1_lcss_vs_ours}
\end{table}

\subsubsection{Human Ratings of Large Models}
Apart from the quantitative evaluation of models on the validation set, we conducted a human evaluation of the best large models for all participants except P01 and P02. We believe it is not suitable to deploy a model in real-world applications based solely on quantitative results without sufficient human conviction in its performance. Hence, we designed an evaluation tool.

The tool consists of a video player displaying participants performing gestures (Fig.~\ref{fig:eval_ui}a), synchronized sensor stream visualizations showing the last two seconds of IMU data (Fig.~\ref{fig:eval_ui}b), and the model's predicted gesture label for the current sliding window (Fig.~\ref{fig:eval_ui}c). Evaluators could mark each prediction as correct or incorrect (Fig.~\ref{fig:eval_ui}d) and replay gesture segments (Fig.~\ref{fig:eval_ui}e) as needed. 

Six human raters (authors) evaluated the large models using this tool. Each rater evaluated models for at least six participants on test set videos (unseen by the models). 
We logged all interactions with the tool and computed the model's precision through majority voting across raters' correct/incorrect ratings. We also computed inter-rater agreement using the AC1 score~\cite{gwet2008computing} to confirm the reliability of the evaluation (see Section~\ref{subsec:human_eval}).

% transferred to a GPU server to train the deep learning model for each participant. We 
% To evaluate performance, we developed a human-centered offline evaluation interface. Procedural details and results of this evaluation are reported in Section~\ref{subsec:human_eval}. This evaluation enabled us to assess model behavior through human judgment and iteratively improve recognition before transferring trained models to participants who opted to receive their personalized deep learning model on their own devices.
% Once we have data from 
% Each participant’s data was used to train one personalized model. For the Transformer, training followed a two-stage process: self-supervised pretraining on raw motion data, followed by supervised fine-tuning on a handful of annotated instances per gesture. For the baseline, gesture templates were compared against new inputs using similarity matching. We report results on held-out test data, visualize model embeddings with t-SNE, and validate predictions through human evaluation.

\subsection{Findings: Effectiveness of Semi-Automatic Annotation Pipeline}
\label{subsec:eval_annotation_pipeline}
The semi-automatic approach achieved an $F_1$-score of $0.95$ (IoU threshold = 0.5), with a mean IoU of $0.88$ for correctly matched segments. It reduced annotation time by 67.3\%, requiring $10.22$ (SD = $5.28$) seconds per gesture compared to $30.46$ (SD = $20.86$) seconds for manual annotation. This represents a $3.0\times$ speedup.
These results demonstrate that the semi-automatic pipeline significantly reduces annotation time while maintaining high overlap with ground truth.
% , while the semi-automatic pipeline detected 224 instances.
% We measured annotation time for both approaches and compared their efficiency, and measured segmentation quality of the semi-automatic approach. 
% To assess segmentation quality, we calculated the Intersection over Union (IoU) between manual (ground truth) and automatic annotations. 

\begin{figure*}[t!]
    \centering
    \includegraphics[width=0.99\textwidth]{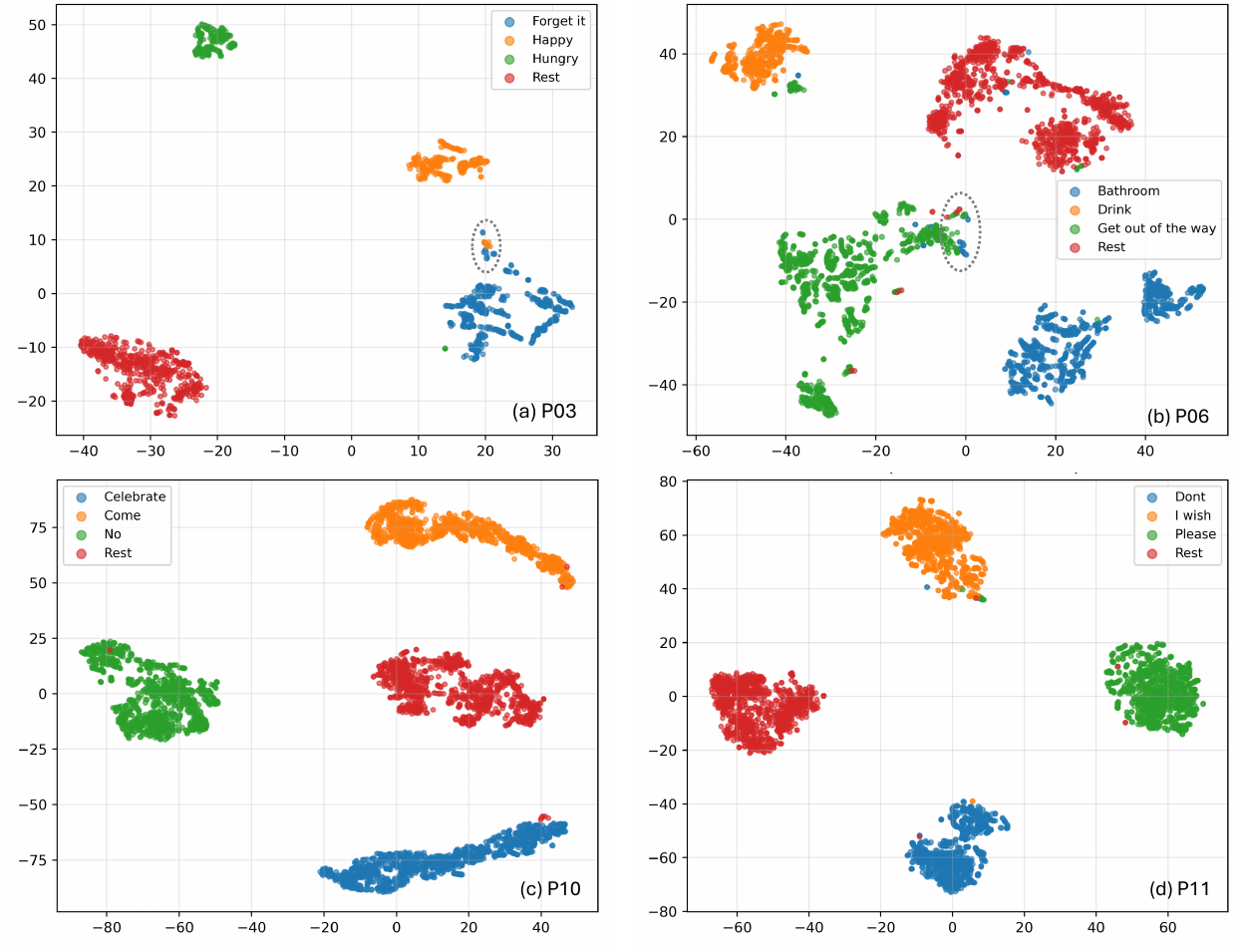}
    \Description{Four t-SNE scatter plot visualizations of gesture embeddings. Panels (a) and (b) show embeddings for motor-impaired participants P03 and P06, respectively, with overlapping clusters highlighted by dotted gray ellipses indicating reduced gesture separability. Panels (c) and (d) show embeddings for non-disabled participants P10 and P11, displaying well-separated gesture clusters.}
    \caption{t-SNE visualizations of gesture embeddings for four participants: (a) P03 and (b) P06 (motor-impaired), and (c) P10 and (d) P11 (non-disabled). Visualizations were generated using each participant's best model according to the $F_1$-score on the validation set. Among motor-impaired users, the model trained for P03 achieved the highest $F_1$-score and P06 the lowest; among non-disabled users, P11 had the lowest $F_1$-score, while P10 achieved near-perfect performance ($F_1 \approx 0.98$). Even the best models for motor-impaired participants (a,b) show reduced gesture separability with overlapping clusters (highlighted by dotted gray ellipses), whereas non-disabled participants (c,d) maintain well-separated gesture clusters even for their weakest models.}
    \label{fig:tsne_plot_model_feat}
\end{figure*}

\subsection{Findings: Effectiveness of Pretraining Strategies}
\label{subsec:finding_pretraining}

Across participants, different pretraining strategies excelled depending on gesture characteristics. 
%
% \rx{Contrastive learning worked best for globally distinct gestures, CPC for long trajectories, and masked reconstruction for gestures distinguished by fine-grained local cues. This suggests personalization must consider not only the user but also the nature of their gesture set.}
%
Below, we discuss the types of gestures for which each strategy performed best.

\textbf{Contrastive learning} (described in Section~\ref{subsubsec:contrastive}) performed best when participants produced gestures that were globally distinct but shared local motor similarities. For example, P08's gestures ``I have a question'' and ``One moment'' involve similar wrist movements but differ in finger articulation, where contrastive learning achieved $F_1 = 0.736$. Similarly, P04's gestures ``Hi'' and ``Look at that'' both begin by lifting the hand from rest but differ only in final positioning (more horizontal for ``Look at that''), with $F_1 = 0.812$.

\textbf{CPC} (described in Section~\ref{subsubsec:cpc}) was most effective for gestures with longer trajectories, as it captures long-range temporal dynamics. P05's gestures ``B,'' ``P,'' and ``R'' all begin with a similar ``P''-shaped movement, differing only in their endings: B adds a half-circle, R adds a downward-right stroke. Distinguishing these required capturing the full temporal trajectory, where CPC achieved $F_1 = 0.82$.

\begin{figure*}[!t]
    \centering
    \includegraphics[width=0.85\textwidth]{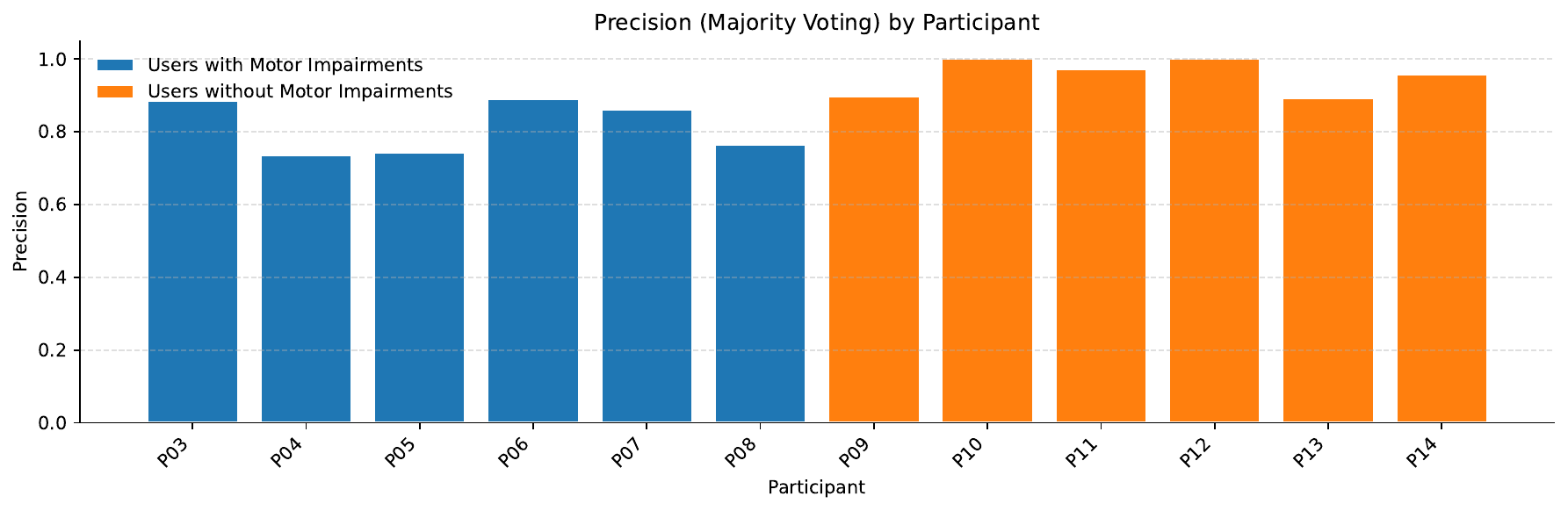}
    \Description{A bar chart showing precision scores computed against majority-voted human ground truth for each participant's best large model. Blue bars represent motor-impaired participants and orange bars represent non-disabled participants. The x-axis shows participant IDs (P03–P14) and the y-axis shows precision values.}
    \caption{Precision of the best large model for each participant (P03–P14) computed using majority-voted ground truth from six independent human raters. This reflects model performance relative to human judgment. The bars with the blue color represent the precision for users with motor impairments, while the bars in orange color represent users without any motor impairment.}
    \label{fig:models_human_rating}
\end{figure*}

\textbf{Masked reconstruction} (described in Section~\ref{subsubsec:masked_recon}) excelled when local motor patterns distinguished gestures, as it trains the model to infer missing segments from the surrounding context. For P03, ``Happy'' and ``Forget it'' appear globally similar but differ in localized motion details, enabling masked reconstruction to achieve $F_1 = 0.929$. Similarly, P06's ``Get out of the way'' and ``Drink'' both involve overlapping wrist movements initially but differ in subtle motion patterns ($F_1 = 0.617$).

\subsection{Findings: Introspection of Large Models}
\label{subsec:introspection}
Motor-impaired participants (P01--P08) had a lower mean $F_1$-score (0.796) compared to non-disabled participants (0.972). t-SNE plots (Fig.~\ref{fig:tsne_plot_model_feat}a and~\ref{fig:tsne_plot_model_feat}b) show that their learned gesture embeddings often overlapped, reflecting partial similarity between gestures. Non-disabled users' gestures, by contrast, were cleanly separable (Fig.~\ref{fig:tsne_plot_model_feat}c and~\ref{fig:tsne_plot_model_feat}d). 
% This also highlights that the system designed for motor-impaired users generalizes well to non-disabled users.

\subsection{Findings: Classification Performance of Baseline and Large Models}
\label{subsec:finding_classification}
% We report results in three parts: (1) comparison of baseline and Transformer models, (2) impact of pretraining strategies, and (3) analysis of gesture complexity and category overlap.
% \subsubsection{Baseline vs. Transformer Models}
Table~\ref{tab:f1_lcss_vs_ours} shows $F_1$-scores across participants. The baseline achieved a mean of $0.589$, while the large model achieved $0.871$. The largest improvements were for motor-impaired users, e.g., P05 ($0.215 \rightarrow 0.820$), P03 ($0.535 \rightarrow 0.929$), and P08 ($0.433 \rightarrow 0.736$). For highly consistent gestures (e.g., P12), the baseline performed competitively (1.00 vs. 0.99 for the large model).
% , even outperforming the Transformer (1.00 vs. 0.99). 
This highlights the trade-off: rule-based models can work for simple, well-separated gestures but fail on complex or overlapping movements, where deep models generalize better.

Although the t-SNE visualization for P10 (Fig.~\ref{fig:tsne_plot_model_feat}c) shows a clear distinction between gesture categories, the $F_1$-score of the baseline model is low. This low score can be attributed to the number of false positives in the predictions. The baseline model misclassified many instances of rest gestures with slight hand movements (e.g., using a phone, typing messages, and writing on paper) as communicative gestures (e.g., no, come, or celebrate). 
% However, performance was good ($F_1$ score over 0.9) for communicative gesture categories.
Additionally, Fig.~\ref{fig:tsne_plot_model_feat} shows the learned embeddings from the large model.
% , which learns to distinguish between gesture categories. 
In contrast, the baseline does not learn representations and relies on template matching. 
% If sensor placement, orientation, or posture differs slightly during inference from that during recording, the baseline model may fail.
% The large model is robust to these slight variations by learning temporal dependencies and gesture trajectories.

\subsection{Findings: Human Ratings of Large Models}
\label{subsec:human_eval}
Human evaluation confirmed the trends observed in Section~\ref{subsec:introspection}: models for motor-impaired users achieved lower mean precision ($0.813$) than those for able-bodied users ($0.953$). Fig.~\ref{fig:models_human_rating} shows the precision of the model by participant. 
The mean precision score for large models across all participants based on human rating was $0.88$, which demonstrates reasonably good alignment between model predictions and human judgments.
This alignment with human judgment suggests the models capture meaningful movement patterns even in challenging scenarios.

Agreement across raters was high (AC1 = $0.92$), indicating that humans were able to consistently interpret gestures and evaluate model behavior. This evaluation also demonstrates the potential of using large models for training new caregivers. The high precision and agreement scores led us to deploy AllyAAC with the large model on our community advisors' mobile devices, where it is currently being used by all three advisors to complement their primary AAC devices.

\section{Discussion and Implications}
\label{sec:discussion}

% \paragraph{\textbf{Current System Performance and Limitations}}
% Our LCSS-based gesture recognition system demonstrates effective performance with a limited number of gesture categories (2-3). However, several challenges emerge as the number of categories increases. The system's prediction accuracy is significantly affected by sensor orientation variations, such as differences in wrist placement (inner versus outer wrist) or postural changes (seated versus standing). These variations introduce inconsistencies in sensor readings due to altered gravitational forces and rotational differences.

% Largely, motor movement patterns are described by their inherent qualities, such as spastic, ataxic, or athetoid. An additional subset of individuals with motor impairment may have minimal movement patterns, such as those associated with neurodegenerative diseases like ALS, or movement characterized by tremor (e.g., Parkinson's). While our algorithm worked consistently well with some movement patterns, it became clear that additional development is needed to consistently and accurately interpret the wide range of movement patterns individuals may have.

\paragraph{\textbf{Gesture Recognition Works but Up to a Point}}
Our current self-attention-based recognizer performed reliably for distinct gestures. But as the gesture vocabulary increased, recognition accuracy declined, particularly when gestures overlapped in motion or intent. In real-world use, the problem is not only technical but conceptual: gestures are situated, embodied, and variable. A flick of the wrist means something different depending on body posture, seating position, or surrounding activity. Our system often failed to account for these contextual shifts.

This limitation is compounded by individual motor profiles. Motor patterns associated with spasticity, fluctuating tone, tremor, or limited range of motion, as seen in people with CP, ALS, or Parkinson’s, can radically alter how gestures manifest. While the system handled some of these patterns well, others exposed the brittleness of our model. Designing for communicative gestures requires moving beyond assumptions of consistent, repeatable motion.

\paragraph{\textbf{Multimodal Models Are Not Just Redundant, They are Necessary}}
While IMU data is lightweight and portable, it has limitations when used alone, particularly when gestures are subtle or sensor placement varies. Our dataset includes synchronized video, offering an opportunity to augment gesture detection with visual cues. Prior work in single-shot gesture customization~\cite{shahi2024visionsingle} and cross-modal learning~\cite{kim2023cross} suggests that combining IMU and vision data can yield stronger, more robust models.

Future systems should treat multimodality not as a fallback but as a design principle: by fusing inertial, visual, and positional signals, models can be more adaptive to individual movement styles and less reliant on narrowly defined motion signatures.

\paragraph{\textbf{From Hand-Tuning to Hidden Complexity}}
One tension that surfaced during user testing was the trade-off between transparency and automation. Participants enjoyed gesture annotation, seeing their own data, but disliked the cumbersome process, such as data transfer for model building and importing the model for real-time deployment. They wanted the system to “just work.” This reflects a broader HCI insight, end-user programming~\cite{ko2004six}. Prior work shows that end-user programming is not easy for general users due to the need for algorithmic thinking and problem decomposition skills. Thus, systems like AllyAAC should make space for user agency in areas that are meaningful (e.g., labeling gestures), while abstracting away the process associated with model training. 
% low-level parameter tuning.

% In future versions of AllyAAC, model creation should be treated as an internal process. Users should be able to annotate gestures and test outcomes, without having to interact with similarity matrices or assign feature weights. This shift will reduce cognitive load and better serve users with varying tech fluency.

\paragraph{\textbf{Learning from Limited Data}}
Many AAC users may only be able or willing to record a few examples of each gesture. This creates a bottleneck for model accuracy. Self-supervised learning offers a way around this. By pretraining on large amounts of unlabeled sensor data—such as individuals' whole recordings (including rest and other daily activities which are not in the gesture set) in our dataset, models learned meaningful motion representations, such as individuals' body language, without requiring exhaustive annotation.

Using techniques such as contrastive learning~\cite{tang2020exploring}, CPC~\cite{haresamudram2021contrastive}, and masked reconstruction~\cite{haresamudram2020masked}, the model was able to learn robust gesture embeddings from raw IMU data. After fine-tuning with just a handful of user-labeled examples, it recognized gestures reasonably well.

\paragraph{\textbf{Designing for the Edges, Not the Center}}
Current gesture recognition benchmarks often rely on data from able-bodied users performing standardized motions in lab conditions. In contrast, our dataset focuses on users with diverse motor capabilities performing idiosyncratic, functional gestures. This reframing is essential: systems built on normative assumptions will always marginalize users with non-normative bodies.

Our dataset can serve as a foundation for reshaping gesture recognition benchmarks. It challenges model developers to handle messier, real-world inputs—and to design systems that adapt to their users, rather than the reverse.

\paragraph{\textbf{Toward Inclusive, Configurable Communication Systems}}
Looking ahead, we envision systems that can adapt to an individual’s movement repertoire over time. Participants proposed features that push this vision forward: gesture sequences for multi-part messages, gesture repetition to signal emphasis, and context-aware triggering based on location. These are not simply interface tweaks; they reflect a desire for expressive range and adaptive control. Designing for AAC means thinking beyond recognition accuracy. It means building systems that preserve autonomy, respect fatigue, and support improvisation.

\paragraph{\textbf{Future Work}}
Our study highlights several avenues for extending AllyAAC. First, while we addressed personalization through self-supervised pretraining and fine-tuning, future work could explore cross-domain transfer strategies to adapt models to new users or sensor placements with minimal labeled data. For example, Thukral et al.\ proposed a \textit{Cross-Domain HAR} framework based on teacher–student self-training~\cite{thukral2025cross}, where a model pre-trained on a large source dataset is iteratively adapted to a target domain. With the release of our dataset, future research can investigate such transfer methods, using our corpus as a foundation for adapting to novel users or contexts.  

Second, participants expressed interest in distinguishing gestures that rely on subtle variations, such as different finger positions, which are difficult to capture with a single wrist-worn IMU. Addressing this need may require integrating multiple sensing modalities, for example, combining IMU with computer vision techniques such as MediaPipe-based hand pose estimation~\cite{zhang2020mediapipe} or distributing lightweight sensors across multiple body locations. Exploring these multimodal or distributed sensing configurations could improve the disambiguation of fine-grained gestures and expand the expressive potential of combined aided and unaided AAC.  

\section{Conclusion}
In this paper, we investigated how to combine aided and unaided AAC to harness the speed and naturalness of body-based gestures while maintaining the intelligibility of speech-generating devices. Through 18 months of participatory design with AAC users who experience motor impairments, we identified key opportunities and challenges for this combination. Based on these findings, we developed AllyAAC, a real-time gesture recognition system that translates personalized body-based movements into synthesized speech.

Our study highlights that gesture recognition for AAC must be context-flexible, robust to sensor variability, and capable of learning from limited labeled data. Equally, such systems must be usable by people whose motor patterns diverge from assumptions embedded in mainstream recognition models. To advance this goal, we contribute a dataset of over 600,000 multimodal data points featuring atypical gestures and an adaptive model that leverages self-supervision and few labeled examples to reduce annotation burden while maintaining strong performance. Our findings indicate that different pretraining strategies work best for different individuals, underscoring the need for personalized approaches.

We argue that AAC technologies must go beyond accuracy. They must be personalized, empowering, and shaped through participatory design with the communities who rely on them. This orientation ensures that technical advances are not only feasible but also meaningful, usable, and impactful in real-world communication.

% \section{Acknowledgment}
\begin{acks}
We thank anonymous reviewers for their insightful feedback. This
work was supported in part by NSF Grant \#2326406 and two seed grants (one from the Penn State Center for Biodevices and another from Penn State Center for Socially Responsible Artificial Intelligence). Some team members are supported in part by Penn State’s SSRI and the Penn State AAC Leadership project (a doctoral training grant funded by the U.S. Department of Education, grant H325D220021).

\end{acks}

\bibliographystyle{ACM-Reference-Format}
% \bibliography{references}
\bibliography{references, Bibliography, Bibliography2, Bibliography3}

\clearpage

%\appendix

%\renewcommand{\thefigure}{A\arabic{figure}}
%\setcounter{figure}{0}
%
%\renewcommand{\thetable}{A\arabic{table}}
%\setcounter{table}{0}

% \input{sections/appendix}

\end{document}